\documentclass[a4paper,UKenglish, autoref, thm-restate]{article}
\usepackage[margin=1in]{geometry}

\usepackage{tikz}
\usepackage{algorithm}
\usepackage{algpseudocodex}
\usetikzlibrary {arrows.meta}

\usepackage{colortbl}
\usepackage{multirow}
\usepackage{makecell}
\usepackage{subcaption}

\usepackage{xspace}

\usepackage{adjustbox}
\usepackage{booktabs}

\PassOptionsToPackage{hyphens}{url}
\usepackage{hyperref}
\hypersetup{
  colorlinks = true,
  citecolor = blue,
  linkcolor = blue,
  urlcolor = blue
}

\usepackage{amsmath}
\newcommand{\argmax}{\operatornamewithlimits{argmax}}
\usepackage{amsthm}
\newtheorem{definition}{Definition}

\usepackage{cleveref}

\usepackage{authblk}

\author[*,1]{Andrew Haberlandt}
\author[*,1]{Harrison Green}
\author[1]{Marijn J.H. Heule}
\affil[1]{Carnegie Mellon University, Pittsburgh, USA}
\affil[ ]{\texttt{\{ahaberla, harrisog, marijn\}@cmu.edu}}
\date{}

\title{Effective Auxiliary Variables\\ via Structured Reencoding}

\usepackage{color}

\newcommand{\tim}[0]{\textcolor{red}{\textbf{T.O.}}}

\newcommand{\vbase}{{\sc Baseline}\xspace}
\newcommand{\vnorand}{{\sc BVA-orig}\xspace}
\newcommand{\vrand}{{\sc BVA-rand-orig}\xspace}
\newcommand{\vheur}{{\sc BVA-rand-3hop}\xspace}

\newcommand{\hbva}{SBVA\xspace}

\newcommand{\mlit}{L}
\newcommand{\mcls}{P}

\begin{document}

\maketitle

\renewcommand{\thefootnote}{\fnsymbol{footnote}}
\footnotetext[1]{Authors contributed equally.}

\begin{abstract}

Extended resolution shows that auxiliary variables are very powerful in theory. However, attempts to exploit this potential in practice have had limited success. One reasonably effective method in this regard is bounded variable addition (BVA), which automatically reencodes formulas by introducing new variables and eliminating clauses, often significantly reducing formula size. We find motivating examples suggesting that the performance improvement caused by BVA stems not only from this size reduction but also from the introduction of \textit{effective auxiliary variables}. Analyzing specific packing-coloring instances, we discover that BVA is \textit{fragile} with respect to formula randomization, relying on variable order to break ties. With this understanding, we augment BVA with a heuristic for breaking ties in a \textit{structured} way. We evaluate our new preprocessing technique, Structured BVA (SBVA), on more than 29\,000 formulas from previous SAT competitions and show that it is robust to randomization. In a simulated competition setting, our implementation outperforms BVA on both randomized and original formulas, and appears to be well-suited for certain families of formulas. 
\end{abstract}

\section{Introduction}
\label{sec:intro}

Pre-processing techniques that introduce and eliminate auxiliary variables have been shown to be helpful both in theory and practice. Theoretically, auxiliary variables lift the power of solvers from the resolution proof system to Extended Resolution (ER)~\cite{tseitin1968complexity, cook1976short, HAKEN1985297}. In practice, efforts to exploit this full power of ER have had limited success; however, auxiliary variables have been used to reencode formulas in a way that drastically reduces their size~\cite{biere2021preprocessing,Biere_Biere_Heule_vanMaaren_Walsh_2009,manthey2012automated}, often leading to a decreased solve time.
In this work we show that this speedup may not be caused entirely by the reduction in formula size, but by the introduction of certain \textit{effective auxiliary variables}.

A very powerful pre-processing technique is \emph{Bounded Variable Elimination} (BVE)~\cite{een2005effective}. As its name suggests, it eliminates a variable $x$ by resolving each clause containing a literal $x$ with every clause containing literal $\overline x$. Importantly, BVE only performs such an elimination if it helps reduce the formula size (measured as the number of clauses plus the number of variables).
A pre-processing technique that introduces new auxiliary variables is \emph{Bounded Variable Addition} (BVA)~\cite{manthey2012automated}, which is the focus of this article. It is well known that the introduction of auxiliary variables is crucial for many succinct encodings (e.g., the Tseitin transformation~\cite{tseitin1968complexity}, or cardinality constraints~\cite{Sinz_2005,kuvcera2019lower}). Following the intuition of BVE, BVA will only introduce a new variable if it can eliminate a larger number of clauses than it adds.

Auxiliary variables may not only be useful to reduce the size of a formula, but they can also capture some semantic meaning about the underlying problem to encode, as we detail in~\Cref{sec:packing}. As a case study, we consider a recent encoding used by Subercaseaux and Heule for computing the packing chromatic number of the infinite square grid via SAT solving~\cite{subercaseaux2023packing}. In their work, BVA was found to generate auxiliary variables that represented clusters of neighboring vertices of the grid. The encoding resulting from running BVA on a direct encoding of the problem inspired a more efficient encoding, by suggesting the usefulness of having auxiliary variables capturing clusters of vertices. In this paper, we offer new insight into this \emph{``meaningful variables''} phenomenon, which we believe can generalize to other problems as well. Furthermore, even though the reencoding resulting from BVA suggested meaningful new variables for the packing coloring problem, it was not as effective as manually designing a more structured encoding based on some of those variables. We take this as motivation to identify shortcomings of BVA and improve upon its design.

In general, on problems where BVA is effective, the effect tends to be extreme. BVA is able to reduce the number of clauses by a $\times 10$ factor or more, improving solve time by  orders of magnitude. However, in this paper, we find that this reduction in solve time is highly sensitive to randomly scrambling the formula (even when controlling for how CDCL solvers are generally sensitive to this form of randomization~\cite{biere2019effect}). In particular, randomizing the order of variables and clauses prior to BVA substantially reduces the positive effect of BVA on solve time, despite maintaining the same overall reduction in formula size. Using the packing $k$-color problem, we show that the effectiveness of BVA relies on the introduction of a few specific variables that account for only a small fraction of the reduction in formula size. Moreover, we identify that the lack of effective tie-breaking in BVA is the cause of this high sensitivity to randomization.
Inspired by these new insights into the behavior of BVA, we present \hbva (\Cref{sec:heuristic}), a version augmented with a tie-breaking heuristic that enables it to introduce better auxiliary variables at each step, even when the original formula is randomized. Our heuristic is based on a connectivity measure between variables in the \emph{incidence graph} of CNF formulas, which is preserved under randomization of the formula. As a result, \hbva is able to identify effective auxiliary variables even when the original formula is scrambled. We evaluate our implementation by running it on more than 29\,000 formulas from the Global Benchmark Database \cite{sat_benchmark}. Experimental results, presented in~\Cref{sec:evaluation}, demonstrate that our approach outperforms the original implementation of BVA.

In summary, the main contributions of this article are:
\begin{enumerate}
	\item We offer new insight into the behavior of BVA, by exhibiting its ability to introduce effective auxiliary variables and showing its sensitivity to formula randomization.
    \item We design \hbva, a heuristic-guided form of BVA, that introduces new variables in a way that is robust to randomization.
    \item We perform a large-scale evaluation of both BVA and \hbva on benchmark problems from the SAT Competition and study their behavior on different families of instances.
    \item We release an open-source implementation of \hbva that supports \textsf{DRAT} proof logging.
\end{enumerate}

\section{Preliminaries}
\label{sec:preliminaries}

A \textit{literal} is either a variable $x$, or its negation ($\overline{x}$). A \emph{propositional formula} in \textit{conjunctive normal form} (CNF) is a conjunction of \textit{clauses}, which are themselves disjunctions of literals. An \textit{assignment} is a mapping from variables to truth values. A positive (negative) literal is true if the corresponding variable is assigned to true (false, respectively). An assignment satisfies a clause if at least one of its literals is true, and we say an \textit{formula} is satisfied if all of its clauses are. A formula is \textit{satisfiable} (SAT) if there exists an assignment that satisfies it, or \textit{unsatisfiable} (UNSAT) otherwise.
For example, the formula $(x \vee \overline{y}) \wedge (\overline{x} \vee z)$ is made up of two clauses, $(x \vee \overline{y})$ and $(\overline{x} \vee z)$, each with two literals. This formula is satisfiable, since the assignment of $x$ and $z$ to true and $y$ to false satisfies it.

\medskip
\noindent\textbf{\textsf{Auxiliary Variables.}} There can be many equivalent ways of encoding a problem into CNF, differing in the meaning assigned to individual variables. Problems often have a \textit{direct encoding}, in which variables are assigned for each individual decision element present in a problem. For example, in a direct encoding of graph coloring, there are $k|V|$ variables, where each $v_{i, c}$ represents whether node $i$ has color $c$ and $k$ is the number of colors.

Although direct encodings are often the most intuitive, more efficient encodings are known for a wide variety of problems. These encodings often add \textit{auxiliary variables} to the formula, which capture properties about a group of variables. One of the simplest examples is an $\mathrm{AtMostOne}(x_1, \dots, x_n)$ constraint, which requires that at most one of the variables $x_1, \dots, x_n$ is true. Without adding auxiliary variables, this constraint requires $\Theta(n^2)$ clauses, which are typically binary clauses between every pair of variables \cite{kuvcera2019lower}. However, with the introduction of auxiliary variables, this constraint can be encoded in a linear number of clauses and variables as follows~\cite{commander}:
\begin{equation}
    \mathrm{AtMostOne}(x_1, \dots, x_n) = \mathrm{AtMostOne}(x_1, x_2, x_3, y) \land \mathrm{AtMostOne}(x_{4}, \dots, x_n, \overline{y})
\end{equation}
where the pairwise encoding is used for $\mathrm{AtMostOne}(x_1, \ldots, x_n)$ where $n < 4$. The split $\mathrm{AtMostOne}$ constraints require that at most one of $\{x_1, x_2, x_3\}$ is true, and at most one of $\{x_4, \dots, x_n\}$ is true, respectively. The added auxiliary variable $y$ prevents a variable in both of the groups from being true. The auxiliary variable $y$ is forced false if any of $x_1$, $x_2$, or $x_3$ are true, and forced true if any of $x_4, \dots, x_n$ are false. If a literal from both groups is true, the auxiliary variable $y$ prevents the formula from being satisfiable.

\medskip
\noindent\textbf{\textsf{Extended Resolution.}} Starting from the original formula, the Extended Resolution proof allows only two simple rules:

\begin{enumerate}
\item \textit{Resolution}: Given clauses $C \vee p$ and $D \vee \overline{p}$, add the clause $C \vee D$ to the proof.
\item \textit{Extension}: Define a new variable $x$ as $x \leftrightarrow \overline{a} \vee \overline{b}$, where $a$ and $b$ are literals in the current proof. Add the clauses $x \vee a$, $x \vee b$, and $\overline{x} \vee \overline{a} \vee \overline{b}$ to the proof.
\end{enumerate}

In \textit{resolution}, the clause $C \vee D$ is implied by the first two clauses, resulting in a logically equivalent formula. In \textit{extension}, however, the introduction of a new variable $x$ is not implied by the original clauses, and results in a formula that preserves satisfiability and is only logically equivalent over the original variables.

Using the extension rule, new variables can be defined in terms of existing variables. The original rule defined by Tseitin \cite{tseitin1968complexity} only allows for definitions of the form $x \leftrightarrow \overline{a} \vee \overline{b}$, the construction for which is given in the definition above. However, the extension rule can be applied repeatedly to construct variables corresponding to arbitrary propositional formulas over the original variables. This flexibility is key to the success of Extended Resolution, but it provides no guidance on how these extensions should be chosen.

\medskip
\noindent\textbf{\textsf{Bounded Variable Addition.}} Bounded Variable Addition (BVA) \cite{manthey2012automated} is a pre-processing technique that reduces the number of clauses in a formula by adding new variables. Each application of BVA first identifies a ``grid'' of clauses, as shown in \autoref{fig:fig-bva}. Then, BVA adds a new variable and clauses which resolve together to generate all clauses in the grid.

\newcommand{\numbercirclebullet}[3]
{
\node[circle,draw,inner sep=1pt,color=black,fill=black,text=white,anchor=center,outer sep=0pt] (lol) at (3.5, #3) {\small{#1}};
\node[anchor=west] (lol) at (3.8, #3) {#2};
}

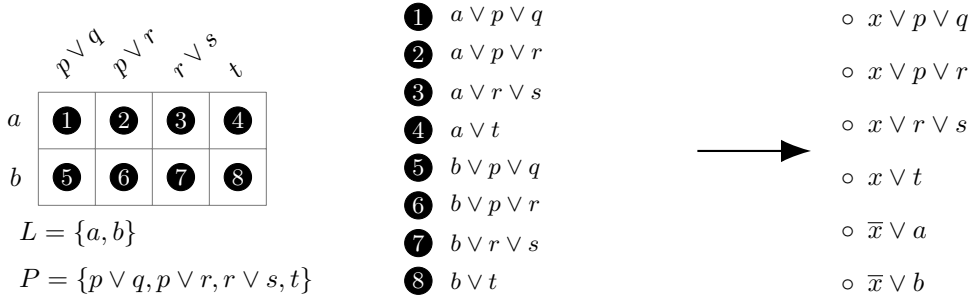
\begin{figure}[h]
    \centering
    \parbox{0.5\textwidth}{
    \centering
    \scalebox{1.0}{\begin{tikzpicture}[]
\draw[step=0.75cm,color=gray] (-1.5,0) grid (1.5,1.5);
\node [rotate=45, anchor=south west] at (-1.125,+1.5) {$p \lor q$};
\node [rotate=45, anchor=south west] at (-0.375,+1.5) {$p \lor r$};
\node [rotate=45, anchor=south west] at (0.375,+1.5) {$r \lor s$};
\node [rotate=45, anchor=south west] at (1.125,+1.5) {$t$};
\node [anchor=east] at (-1.6, +1.125) {$a$};
\node [anchor=east] at (-1.6, +0.375) {$b$};

\node[circle,draw,inner sep=1pt,color=black,fill=black,text=white] (lol) at (-1.125,+1.125) {\small{1}};
\node[circle,draw,inner sep=1pt,color=black,fill=black,text=white] (lol) at (-0.375,+1.125) {\small{2}};
\node[circle,draw,inner sep=1pt,color=black,fill=black,text=white] (lol) at (+0.375,+1.125) {\small{3}};
\node[circle,draw,inner sep=1pt,color=black,fill=black,text=white] (lol) at (+1.125,+1.125) {\small{4}};

\node[circle,draw,inner sep=1pt,color=black,fill=black,text=white] (lol) at (-1.125,+0.375) {\small{5}};
\node[circle,draw,inner sep=1pt,color=black,fill=black,text=white] (lol) at (-0.375,+0.375) {\small{6}};
\node[circle,draw,inner sep=1pt,color=black,fill=black,text=white] (lol) at (+0.375,+0.375) {\small{7}};
\node[circle,draw,inner sep=1pt,color=black,fill=black,text=white] (lol) at (+1.125,+0.375) {\small{8}};

\node[anchor=west] at (-1.875, -0.375) {$\mlit = \{a, b\}$};
\node[anchor=west] at (-1.875, -1.0) {$\mcls = \{p \lor q, p \lor r, r \lor s, t\}$};

\numbercirclebullet{1}{\small{$a \lor p \lor q$}}{2.5};
\numbercirclebullet{2}{\small{$a \lor p \lor r$}}{2.0};
\numbercirclebullet{3}{\small{$a \lor r \lor s$}}{1.5};
\numbercirclebullet{4}{\small{$a \lor t$}}{1.0};
\numbercirclebullet{5}{\small{$b \lor p \lor q$}}{0.5};
\numbercirclebullet{6}{\small{$b \lor p \lor r$}}{0.0};
\numbercirclebullet{7}{\small{$b \lor r \lor s$}}{-0.5};
\numbercirclebullet{8}{\small{$b \lor t$}}{-1.0};

\end{tikzpicture}
}}
\hspace{0.5in}
\parbox{0.2in}{
\begin{tikzpicture}
\draw [-{Latex[length=4mm]}, thick] (-0.5, 0) -- (1.0, 0.0);
\end{tikzpicture}
}
\hspace{0.25in}
\parbox{0.2\textwidth}{
\begin{itemize}
    \item[$\circ$] $x \lor p \lor q$
    \item[$\circ$] $x \lor p \lor r$
    \item[$\circ$] $x \lor r \lor s$
    \item[$\circ$] $x \lor t$
    \item[$\circ$] $\overline{x} \lor a$
    \item[$\circ$] $\overline{x} \lor b$
\end{itemize}
}
\caption{Bounded variable addition transforms groups of clauses (those that form a grid) by adding a new variable and eliminating a number of clauses.}
    \label{fig:fig-bva}
\end{figure} 
Collectively, for a formula $F$, the grid constitutes a set of literals $\mlit$ and a set of partial clauses $\mcls$, such that $ \forall l \in \mlit, \forall C \in \mcls : (l \vee C) \in F$. While bounded-variable elimination eliminates variables by replacing all the clauses containing a variable with their resolvents, BVA tries to identify grids of \textit{resolvents} which can be generated by the introduction of a new variable and a smaller number of clauses. These grids of clauses capture the fact that either the entirety of $\mlit$ must be satisfied, or the entirety of $\mcls$ must be satisfied. More precisely, $F \implies  (\bigwedge_{l \in \mlit}{l}) \lor (\bigwedge_{C \in \mcls}{C})$.
By identifying these grids, BVA replaces $|\mlit| \cdot |\mcls|$ clauses with a single, new variable $x$ and $|\mlit| + |\mcls|$ clauses (which can generate the original set by resolution on $x$ in $\{x \lor C \mid C \in \mcls\} \times \{\overline{x} \lor l \mid l \in \mlit\}$). Therefore, if $|\mlit| \cdot |\mcls| > |\mlit| + |\mcls| + 1$, then this replacement results in a reduction in formula size.

Note that a BVA replacement step can be simulated by extended resolution: First, add the definition $x \leftrightarrow \mathrm{AND}(\mlit)$. In the example
above, this means adding the clauses $x \lor \overline{a} \lor \overline{b}$, $\overline x \lor a$, and $\overline x \lor b$. Afterwards, the clauses
$x \lor p \lor q$, $x \lor p \lor r$, $x \lor r \lor s$, and $x \lor t$ can each be derived using $|\mlit|$ resolution steps. For example, to derive $x \lor p \lor q$, resolve 
$x \lor \overline{a} \lor \overline{b}$ with $a \lor p \lor q$ and the result with $b \lor p \lor q$. Afterward the clauses used in these resolution steps can be deleted. 

\medskip
\noindent\textbf{\textsf{The {\sc SimpleBoundedVariableAddition} algorithm.}} Manthey et al.\ \cite{manthey2012automated} propose a greedy algorithm to identify these grids of resolvents that prioritizes literals which appear in many clauses called {\sc SimpleBoundedVariableAddition}.
An abbreviated version of a single variable addition in this algorithm is shown in \autoref{alg:bva}.

\begin{algorithm}[ht]
\caption{A single variable addition in {\sc SimpleBoundedVariableAddition} \cite{manthey2012automated}} 
\label{alg:bva}
\vspace{0.5em}

\begin{algorithmic}[1]
    \Statex \Call{PartialClauses}{$F$, $l$} := $\{C \setminus \{l\} \mid (C \in F) \land (l \in C)\}$
    \Statex $F := \text{the clauses in the current formula}$
    \Statex $l := \text{a literal in } F$
    \medbreak
    \State $\mlit \gets \{l\}$
    \State $\mcls \gets \Call{PartialClauses}{F, l}$

\While{True}
        \State $l_\mathrm{max} = \argmax_{l_{m} \in \mathrm{Lits}(F)} \left| \mcls \cap \Call{PartialClauses}{F, l_m} \right|$
        \Comment{Sensitive to tiebreaking}
        \If{adding $l_\mathrm{max}$ results in a greater reduction}
        \State $\mlit \gets \mlit \cup \{l_\mathrm{max}\}$
\State $\mcls \gets \mcls \cap \Call{PartialClauses}{F, l_\mathrm{max}}$
        \Else
        \State \textbf{break}
        \EndIf
    \EndWhile{}
\If{$|\mlit| \cdot |\mcls| > |\mlit| + |\mcls| + 1$}
    \Comment{If adding this variable would reduce the formula size}
    \State $S_\mathrm{add} \gets \{ x \lor C \mid C \in \mcls\} \cup \{ \overline{x} \lor l_{m} \mid l_{m} \in \mlit\}$
    \Comment{Introduce a new variable $x$}
    \State $S_\mathrm{remove} \gets \{ l_{i} \lor C \mid (l_{i}, C) \in \mlit \times \mcls\}$
    \State $F \gets (F \setminus S_\mathrm{remove}) \cup S_\mathrm{add}$
    \EndIf
\end{algorithmic}
\end{algorithm}

Each identified grid starts from the most frequently occurring literal $l$ in the current formula. The grid starts with dimension $1 \times \left| F_{l} \right|$, where $F_l$ is the set of clauses containing $l$. From there, the algorithm searches for a literal $l_\mathrm{max}$ to add to the grid, which maximizes the number of remaining resolvents.

\begin{figure}[h]
\centering
\parbox{0.92\textwidth}{
\centering
\parbox{0.36\textwidth}{\scalebox{0.8}{
\begin{tikzpicture}[]
\draw[step=0.75cm,color=gray] (0,0) grid (3.0,-2.25);
\node [rotate=45, anchor=south west] at (0.375,0.0) {$p \lor q$};
\node [rotate=45, anchor=south west] at (1.125,0.0) {$p \lor r$};
\node [rotate=45, anchor=south west] at (1.875,0.0) {$s \lor t$};
\node [rotate=45, anchor=south west] at (2.625,0.0) {$u$};
\node[mark size=2pt,color=black] at (0.375,-0.375) {\pgfuseplotmark{*}};
\node[mark size=2pt,color=black] at (1.125,-0.375) {\pgfuseplotmark{*}};
\node[mark size=2pt,color=black] at (1.875,-0.375) {\pgfuseplotmark{*}};
\node[mark size=2pt,color=black] at (2.625,-0.375) {\pgfuseplotmark{*}};
\node[mark size=2pt,color=black] at (0.375,-1.125) {\pgfuseplotmark{*}};
\node[mark size=2pt,color=black] at (1.125,-1.125) {\pgfuseplotmark{*}};
\node[mark size=2pt,color=black] at (1.875,-1.125) {\pgfuseplotmark{*}};
\node[mark size=2pt,color=black] at (1.125,-1.875) {\pgfuseplotmark{*}};
\node [anchor=east] at (-0.1499999999999999,-0.375) {$a$};
\node [anchor=east] at (-0.1499999999999999,-1.125) {$b$};
\node [anchor=east] at (-0.1499999999999999,-1.875) {$c$};
\node[rectangle, draw=blue, line width=0.4mm, minimum width=3.0cm, minimum height=0.75cm, anchor=north west] at (0.0,0.0) {};
\node [anchor=west] at (0.0,-2.625) {$\mlit = \{a\}$};
\node [anchor=west] at (0.0,-3.125) {$\mcls = \{p \lor q, p \lor r, s \lor t, u\}$};
\node [anchor=west] at (0.0,-3.625) {$R = 4 - 5 = -1$};
\draw [-{Latex[length=4mm]}, thick] (3.375,-1.125) -- (6.4,-1.125);
\end{tikzpicture}
}}
\parbox{0.33\textwidth}{\scalebox{0.8}{
\begin{tikzpicture}[]
\draw[step=0.75cm,color=gray] (0,0) grid (3.0,-2.25);
\node [rotate=45, anchor=south west] at (0.375,0.0) {$p \lor q$};
\node [rotate=45, anchor=south west] at (1.125,0.0) {$p \lor r$};
\node [rotate=45, anchor=south west] at (1.875,0.0) {$s \lor t$};
\node [rotate=45, anchor=south west] at (2.625,0.0) {$u$};
\node[mark size=2pt,color=black] at (0.375,-0.375) {\pgfuseplotmark{*}};
\node[mark size=2pt,color=black] at (1.125,-0.375) {\pgfuseplotmark{*}};
\node[mark size=2pt,color=black] at (1.875,-0.375) {\pgfuseplotmark{*}};
\node[mark size=2pt,color=black] at (2.625,-0.375) {\pgfuseplotmark{*}};
\node[mark size=2pt,color=black] at (0.375,-1.125) {\pgfuseplotmark{*}};
\node[mark size=2pt,color=black] at (1.125,-1.125) {\pgfuseplotmark{*}};
\node[mark size=2pt,color=black] at (1.875,-1.125) {\pgfuseplotmark{*}};
\node[mark size=2pt,color=black] at (1.125,-1.875) {\pgfuseplotmark{*}};
\node[rectangle, draw=blue, line width=0.4mm, minimum width=2.25cm, minimum height=1.5cm, anchor=north west] at (0.0,0.0) {};
\node [anchor=west] at (0.0,-2.625) {$\mlit = \{a, b\}$};
\node [anchor=west] at (0.0,-3.125) {$\mcls = \{p \lor q, p \lor r, s \lor t\}$};
\node [anchor=west] at (0.0,-3.625) {$R = 6 - 5 = 1$};
\draw [-{Latex[length=4mm]}, thick] (3.375,-1.125) -- (6.5,-1.125);
\end{tikzpicture}
}}
\parbox{0.21\textwidth}{\scalebox{0.8}{
\begin{tikzpicture}[]
\draw[step=0.75cm,color=gray] (0,0) grid (3.0,-2.25);
\node [rotate=45, anchor=south west] at (0.375,0.0) {$p \lor q$};
\node [rotate=45, anchor=south west] at (1.125,0.0) {$p \lor r$};
\node [rotate=45, anchor=south west] at (1.875,0.0) {$s \lor t$};
\node [rotate=45, anchor=south west] at (2.625,0.0) {$u$};
\node[mark size=2pt,color=black] at (0.375,-0.375) {\pgfuseplotmark{*}};
\node[mark size=2pt,color=black] at (1.125,-0.375) {\pgfuseplotmark{*}};
\node[mark size=2pt,color=black] at (1.875,-0.375) {\pgfuseplotmark{*}};
\node[mark size=2pt,color=black] at (2.625,-0.375) {\pgfuseplotmark{*}};
\node[mark size=2pt,color=black] at (0.375,-1.125) {\pgfuseplotmark{*}};
\node[mark size=2pt,color=black] at (1.125,-1.125) {\pgfuseplotmark{*}};
\node[mark size=2pt,color=black] at (1.875,-1.125) {\pgfuseplotmark{*}};
\node[mark size=2pt,color=black] at (1.125,-1.875) {\pgfuseplotmark{*}};
\node[rectangle, draw=red, line width=0.4mm, minimum width=0.75cm, minimum height=2.25cm, anchor=north west] at (0.75,0.0) {};
\node [anchor=west] at (0.0,-2.625) {$\mlit = \{a, b, c\}$};
\node [anchor=west] at (0.0,-3.125) {$\mcls = \{p \lor r\}$};
\node [anchor=west] at (0.0,-3.625) {$R = 3 - 4 = -1$};
\end{tikzpicture}
}}
}
\caption{BVA adds variables to form a grid, until the reduction stops increasing. Here, the largest reduction was 1, and the variable corresponding to the middle grid will be added.}
\label{fig:bva_reduction}
\end{figure}
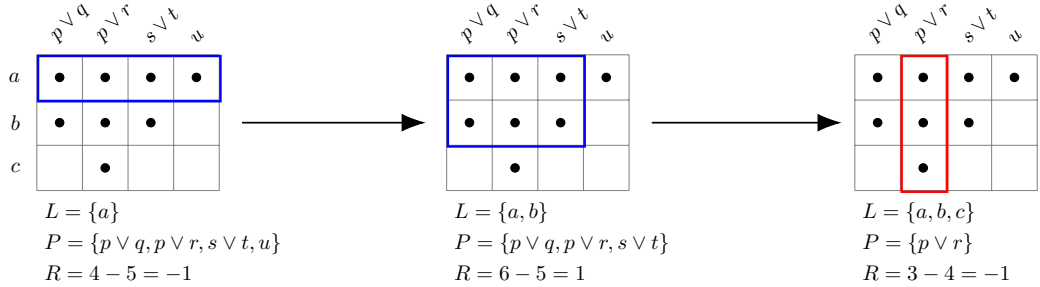

To identify the literal $l_\mathrm{max}$, the BVA algorithm looks for the literal for which $(l_\mathrm{max} \lor C)$ appears in $F$ for the greatest number of clauses $C \in \mcls$ (line 4).
At each step, a literal is added to $\mlit$ (line 6), and clauses may be removed from $\mcls$ (line 7). The grid will continue to shrink until the addition of a literal to the grid would not increase the size of the formula reduction (line 5), as shown in \autoref{fig:bva_reduction}.

In BVA, variable additions are performed as long as there is a reduction in formula size. The entirety of \autoref{alg:bva} is repeated using different literals for $l$ to construct multiple new auxiliary variables. Specifically, the original implementation defines a priority queue of literals ordered by the number of clauses each literal appears in. Our adaptation of BVA (\autoref{sec:heuristic}) reuses this implementation detail. These repeated applications of BVA enable the algorithm to achieve large reductions in formula size, and auxiliary variables added in previous steps can even be re-used in future variable introductions.

\section{Motivating Example}
\label{sec:packing}

To motivate the need for a heuristic-guided version of BVA, we will first demonstrate the effect of randomization on existing implementations of BVA, and the disproportionate impact of a few critical variable additions.

\subsection{Packing Colorings}
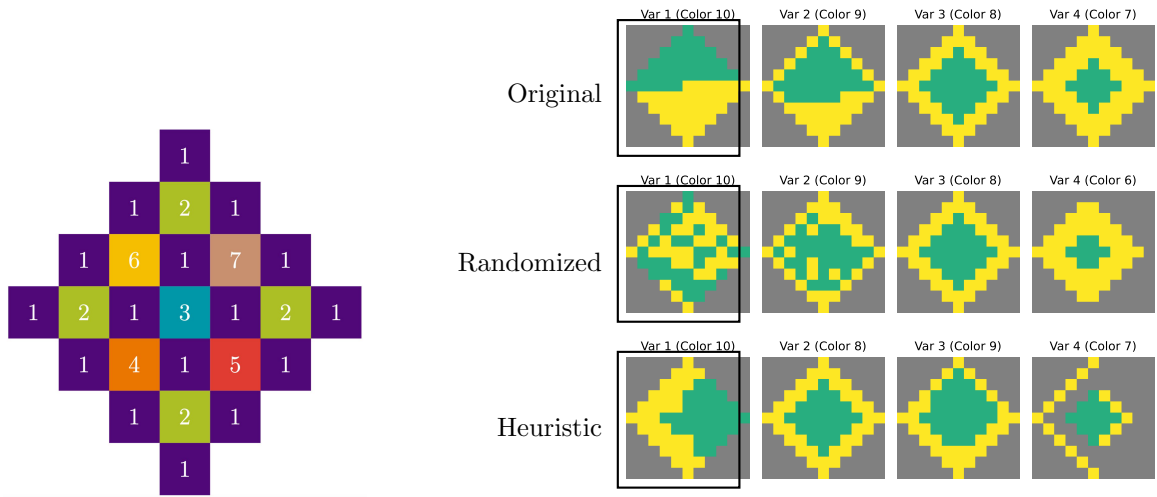
\begin{figure}[!b]
    \centering
    \captionsetup[subfigure]{justification=centering}
    \subcaptionbox{\label{fig:fig-coloring-fig}Figure from \cite{subercaseaux2023packing} showing the $D_{3, 5}$ grid packing k-coloring problem}{\makebox[0.30\textwidth][c]{
        \includegraphics[width=0.30\textwidth]{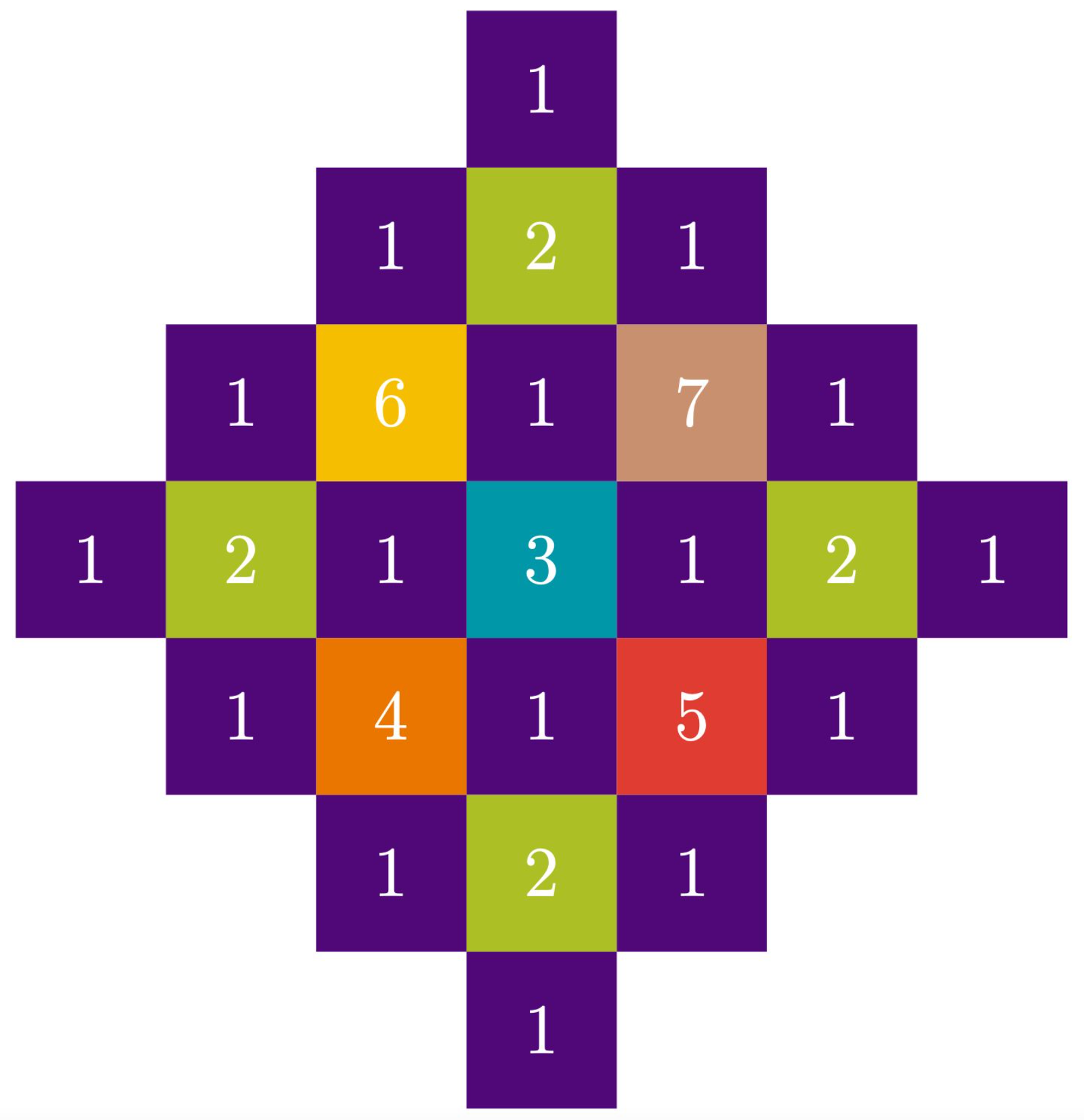}
    }}
    \subcaptionbox{\label{fig:first-four}The effect of variable randomization on the first four BVA substitutions in $D_{5, 10}$. The black boxes indicate the first variable addition, the effect of which is isolated in \autoref{table:table-packing}.\label{fig:fig3b}}{\makebox[0.68\textwidth][c]{
        \begin{tabular}{c}
            \begin{tikzpicture}
                \node[anchor=south west,inner sep=0] at (0,0) {
                    \includegraphics[width=0.45\textwidth]{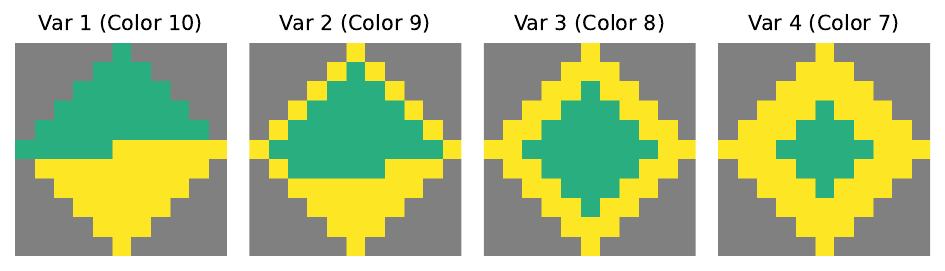}
                };
                \draw[black,thick] (0.0,0) rectangle (1.60,1.8);
                \node[anchor=east,inner sep=0] at (-0.2, 0.8) {Original};
                \node[anchor=east,inner sep=0] at (-3.0, 0) {};
                \node[anchor=east,inner sep=0] at (7.5, 0) {};
            \end{tikzpicture}\\
            \begin{tikzpicture}
                \node[anchor=south west,inner sep=0] at (0.0,0.0) {
                    \includegraphics[width=0.45\textwidth]{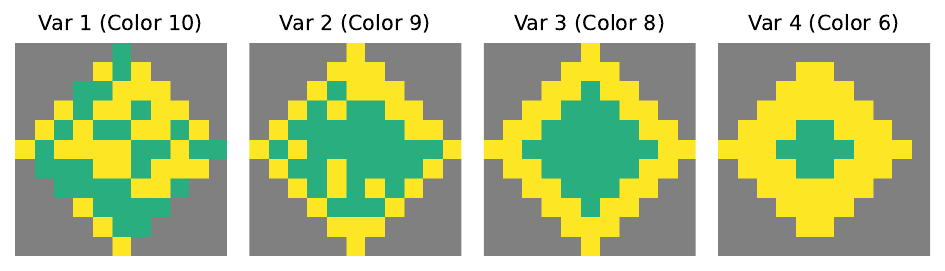}
                };
                \draw[black,thick] (0.0,0) rectangle (1.60,1.8);
                \node[anchor=east,inner sep=0] at (-0.2, 0.8) {Randomized};
                \node[anchor=east,inner sep=0] at (-3.0, 0) {};
                \node[anchor=east,inner sep=0] at (7.5, 0) {};
            \end{tikzpicture}\\
            \begin{tikzpicture}
                \node[anchor=south west,inner sep=0] at (0.0,0.0) {
                    \includegraphics[width=0.45\textwidth]{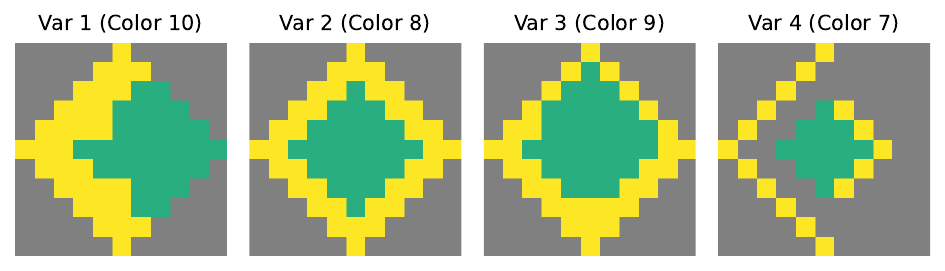}
                };
                \draw[black,thick] (0.0,0) rectangle (1.60,1.8);
                \node[anchor=east,inner sep=0] at (-0.2, 0.8) {Heuristic};
                \node[anchor=east,inner sep=0] at (-3.0, 0) {};
                \node[anchor=east,inner sep=0] at (7.5, 0) {};
            \end{tikzpicture}
        \end{tabular}
    }}
    \caption{The auxiliary variables introduced by BVA on the packing k-coloring problem are sensitive to randomization.}
    \label{fig:fig-randomization}
\end{figure}

BVA has been shown to be effective on the grid packing $k$-coloring problems, whose constraints are based on coloring a circular grid of tiles, shown in \autoref{fig:fig-coloring-fig}. Unlike a standard graph coloring problem, each color in the packing $k$-coloring problem is associated with a integer distance from 1 to $k$. When coloring the grid, two tiles can only have the same color if the taxicab distance between them is greater than the color number. For example, two tiles of color 3 must have at least 3 tiles between them, while color 1 tiles cannot be adjacent. The $D_{r, k}$ problem asks whether the grid of radius $r$ can be colored with $k$ colors. 

The direct encoding for this problem consists of variables $v_{i, c}$, denoting that grid location $i$ has color $c$. There are three types of clauses \cite{subercaseaux2023packing}: 

\begin{enumerate}
\item At-Least-One-Color: $\forall i, (v_{i,1} \vee v_{i,2} \vee \dots \vee v_{i,k})$. Each tile must be colored with a color between 1 and $k$.
\item At-Most-One-Distance: $\forall i, j, c : d(i, j) \leq c, (\overline{v_{i,c}} \vee \overline{v_{j,c}})$. If the distance between two tiles is less than or equal to the color, they cannot both have that color. 
\item Center-Clause: $v_{(0, 0), c}$ for a fixed color $c$. This is a symmetry-breaking optimization \cite{subercaseaux2022packing}, which has no effect on BVA since it ignores unit clauses.
\end{enumerate}

Previous work \cite{subercaseaux2023packing} showed that BVA can reduce the size of such formulas by a factor of $4$, and induces more than a $\times 4$ speedup on the larger instance ($D_{6, 11}$). They found that auxiliary variables capture \textit{regions} of grid tiles within a particular color, i.e. the grid replacement happens entirely within the binary at-most-one-distance constraints.

We visualize the variables introduced by BVA on $D_{5, 10}$, the packing $k$-coloring problem with radius 5 and 10 colors. In the first row of \autoref{fig:first-four}, each of the four plots introduces a new auxiliary variable $x$ for one of the colors $c \in \{1, \dots, 10\}$ (denoted above each plot). All the binary clauses for color $c$ (At-Most-One-Distance clauses) of the form $(\overline{v_{i,c}} \vee \overline{v_{j,c}})$ with grid location $i$ corresponding to a green square and grid location $j$ corresponding to a yellow square will be replaced with a smaller number of clauses: $(x \vee \overline{v_{i,c}})$ for each green location $i$ and $(\overline{x} \vee \overline{v_{j,c}})$ for each yellow location $j$.

\subsection{Negative Impact of Randomization}

We discovered that randomizing packing $k$-coloring formulas prior to running BVA significantly increases the resulting solve time. Furthermore, the variables added by BVA after randomization fail to capture the clustered \textit{regions} within the problem's 2D space that are identified without randomization. \autoref{fig:first-four} shows the first four variable additions performed by BVA on $D_{5, 10}$. The effect is especially noticeable in the first few variable additions.
The structure of these variables is more than a visual artifact. Running BVA to completion produces a formula that requires more than $\times2$ the solve time in CaDiCaL compared with running BVA on the original formula, despite a similar reduction in formula size (see \autoref{table:table-packing}).

We found that the \textit{first} variable added by BVA in $D_{5, 10}$ had a disproportionate impact on the solve time of the formula. We isolated the effect of a single replacement by allowing BVA to only produce one new auxiliary variable and then evaluating the solve time of the resulting formula. \autoref{table:table-packing} shows that a single variable addition (outlined in black in \autoref{fig:first-four}) can achieve a $\times 6$ speedup over the original formula and that the impact of this single addition is also substantially affected by randomization. Although randomization before BVA did not affect the size reduction of the first variable addition, the randomized formula with a single BVA step is $2$ times slower compared to the original formula.

The importance of individual variable additions and their sensitivity to randomization suggests that BVA's impact is derived not only from the size reduction but from the \textit{structure} of the variable additions.

\begin{table}[t]
    \centering
    \caption{CaDiCaL solve time for the $D_{5, 10}$ packing problem, breaking BVA ties using the original variable order (sorted), randomized variable order (randomized), or the heuristic proposed in \autoref{sec:heuristic} (heuristic). Breaking ties differently has a \textit{significant} effect on solve time even when the resulting formula is the same size (Single BVA Step).}
    \begin{tabular}{lrrr}
        \toprule
        & \# Vars & \# Clauses & Solve (s) \\
        \midrule
        Original formula& 610 & 10688 & 590.545 \\
        \midrule
        Single BVA Step (sorted)& 611 & 9819 & 105.635 \\
        Single BVA Step (randomized)& 611 & 9819 & 429.396 \\
        Single BVA Step (heuristic, \textbf{this paper})& 611 & 9819 & 213.018 \\
        \midrule
        Full BVA (sorted)& 973 & 2313 & 38.749 \\
        Full BVA (randomized)& 971 & 2305 & 107.675 \\
        Full BVA (heuristic, \textbf{this paper})& 972 & 2290 & 55.482 \\
        \bottomrule
    \end{tabular}
    \label{table:table-packing}
\end{table}

\subsection{Ties in Bounded Variable Addition}

\begin{figure}[ht]
\parbox{0.5\textwidth}{\hspace{1.00cm}
\vspace{-1.25cm}
\parbox{0.33\textwidth}{\scalebox{0.8}{
\begin{tikzpicture}[]
\draw[step=0.75cm,color=gray] (0,0) grid (3.75,-2.25);
\node [anchor=east] at (-0.1499999999999999,-0.375) {$a$};
\node [anchor=east] at (-0.1499999999999999,-1.125) {$b$};
\node [anchor=east] at (-0.1499999999999999,-1.875) {$c$};
\node [rotate=45, anchor=south west] at (0.375,0.0) {$p \lor q$};
\node [rotate=45, anchor=south west] at (1.125,0.0) {$p \lor r$};
\node [rotate=45, anchor=south west] at (1.875,0.0) {$r \lor s$};
\node [rotate=45, anchor=south west] at (2.625,0.0) {$t$};
\node [rotate=45, anchor=south west] at (3.375,0.0) {$u$};
\node[mark size=2pt,color=black] at (0.375,-0.375) {\pgfuseplotmark{*}};
\node[mark size=2pt,color=black] at (1.125,-0.375) {\pgfuseplotmark{*}};
\node[mark size=2pt,color=black] at (1.875,-0.375) {\pgfuseplotmark{*}};
\node[mark size=2pt,color=black] at (2.625,-0.375) {\pgfuseplotmark{*}};
\node[mark size=2pt,color=black] at (3.375,-0.375) {\pgfuseplotmark{*}};
\node[mark size=2pt,color=black] at (0.375,-1.125) {\pgfuseplotmark{*}};
\node[mark size=2pt,color=black] at (1.125,-1.125) {\pgfuseplotmark{*}};
\node[mark size=2pt,color=black] at (1.875,-1.125) {\pgfuseplotmark{*}};
\node[mark size=2pt,color=black] at (1.125,-1.875) {\pgfuseplotmark{*}};
\node[mark size=2pt,color=black] at (1.875,-1.875) {\pgfuseplotmark{*}};
\node[mark size=2pt,color=black] at (2.625,-1.875) {\pgfuseplotmark{*}};
\node[rectangle, draw=blue, line width=0.4mm, minimum width=3.75cm, minimum height=0.75cm, anchor=north west] at (0.0,0.0) {};
\node [anchor=west] at (0.0,-2.625) {$\mlit = \{a\}$};
\node [anchor=west] at (0.0,-3.125) {$\mcls = \{p \lor q, p \lor r, r \lor s, t, u\}$};
\node [anchor=west] at (0.0,-3.625) {$R = 5 - 6 = -1$};
\draw [-{Latex[length=4mm]}, thick] (4.1, -1.1) -- (7.5, 0.5);
\draw [-{Latex[length=4mm]}, thick] (4.1, -1.1) -- (7.5, -2.5);
\end{tikzpicture}
}}
}\parbox{0.5\textwidth}{\parbox{0.4\textwidth}{\scalebox{0.8}{
\begin{tikzpicture}[]
\draw[step=0.75cm,color=gray] (0,0) grid (3.75,-2.25);
\node [anchor=east] at (-0.1499999999999999,-0.375) {$a$};
\node [anchor=east] at (-0.1499999999999999,-1.125) {$b$};
\node [anchor=east] at (-0.1499999999999999,-1.875) {$c$};
\node [rotate=45, anchor=south west] at (0.375,0.0) {$p \lor q$};
\node [rotate=45, anchor=south west] at (1.125,0.0) {$p \lor r$};
\node [rotate=45, anchor=south west] at (1.875,0.0) {$r \lor s$};
\node [rotate=45, anchor=south west] at (2.625,0.0) {$t$};
\node [rotate=45, anchor=south west] at (3.375,0.0) {$u$};
\node[mark size=2pt,color=black] at (0.375,-0.375) {\pgfuseplotmark{*}};
\node[mark size=2pt,color=black] at (1.125,-0.375) {\pgfuseplotmark{*}};
\node[mark size=2pt,color=black] at (1.875,-0.375) {\pgfuseplotmark{*}};
\node[mark size=2pt,color=black] at (2.625,-0.375) {\pgfuseplotmark{*}};
\node[mark size=2pt,color=black] at (3.375,-0.375) {\pgfuseplotmark{*}};
\node[mark size=2pt,color=black] at (0.375,-1.125) {\pgfuseplotmark{*}};
\node[mark size=2pt,color=black] at (1.125,-1.125) {\pgfuseplotmark{*}};
\node[mark size=2pt,color=black] at (1.875,-1.125) {\pgfuseplotmark{*}};
\node[mark size=2pt,color=black] at (1.125,-1.875) {\pgfuseplotmark{*}};
\node[mark size=2pt,color=black] at (1.875,-1.875) {\pgfuseplotmark{*}};
\node[mark size=2pt,color=black] at (2.625,-1.875) {\pgfuseplotmark{*}};
\node[rectangle, draw=blue, line width=0.4mm, minimum width=2.25cm, minimum height=1.5cm, anchor=north west] at (0.0,0.0) {};
\node [anchor=west] at (3.75,-0.375) {$\mlit = \{a, b\}$};
\node [anchor=west] at (3.75,-0.875) {$\mcls = \{p \lor q, p \lor r, r \lor s\}$};
\node [anchor=west] at (3.75,-1.375) {$R = 6 - 5 = 1$};
\end{tikzpicture}
}}
\parbox{0.4\textwidth}{\scalebox{0.8}{
\begin{tikzpicture}[]
\draw[step=0.75cm,color=gray] (0,0) grid (3.75,-2.25);
\node [anchor=east] at (-0.1499999999999999,-0.375) {$a$};
\node [anchor=east] at (-0.1499999999999999,-1.125) {$b$};
\node [anchor=east] at (-0.1499999999999999,-1.875) {$c$};
\node [rotate=45, anchor=south west] at (0.375,0.0) {$p \lor q$};
\node [rotate=45, anchor=south west] at (1.125,0.0) {$p \lor r$};
\node [rotate=45, anchor=south west] at (1.875,0.0) {$r \lor s$};
\node [rotate=45, anchor=south west] at (2.625,0.0) {$t$};
\node [rotate=45, anchor=south west] at (3.375,0.0) {$u$};
\node[mark size=2pt,color=black] at (0.375,-0.375) {\pgfuseplotmark{*}};
\node[mark size=2pt,color=black] at (1.125,-0.375) {\pgfuseplotmark{*}};
\node[mark size=2pt,color=black] at (1.875,-0.375) {\pgfuseplotmark{*}};
\node[mark size=2pt,color=black] at (2.625,-0.375) {\pgfuseplotmark{*}};
\node[mark size=2pt,color=black] at (3.375,-0.375) {\pgfuseplotmark{*}};
\node[mark size=2pt,color=black] at (0.375,-1.125) {\pgfuseplotmark{*}};
\node[mark size=2pt,color=black] at (1.125,-1.125) {\pgfuseplotmark{*}};
\node[mark size=2pt,color=black] at (1.875,-1.125) {\pgfuseplotmark{*}};
\node[mark size=2pt,color=black] at (1.125,-1.875) {\pgfuseplotmark{*}};
\node[mark size=2pt,color=black] at (1.875,-1.875) {\pgfuseplotmark{*}};
\node[mark size=2pt,color=black] at (2.625,-1.875) {\pgfuseplotmark{*}};
\node[rectangle, draw=blue, line width=0.4mm, minimum width=2.25cm, minimum height=0.75cm, anchor=north west] at (0.75,0.0) {};
\node[rectangle, draw=blue, line width=0.4mm, minimum width=2.25cm, minimum height=0.75cm, anchor=north west] at (0.75,-1.5) {};
\node [anchor=west] at (3.75,-0.375) {$\mlit = \{a, c\}$};
\node [anchor=west] at (3.75,-0.875) {$\mcls = \{p \lor q, p \lor r, r \lor s\}$};
\node [anchor=west] at (3.75,-1.375) {$R = 6 - 5 = 1$};
\end{tikzpicture}
}}
}\caption{The addition of $b$ or $c$ both lead to a $2 \times 3$ grid of resolvents. Breaking this tie in different ways leads to different variable additions.}
\label{fig:bva_ties}
\end{figure}
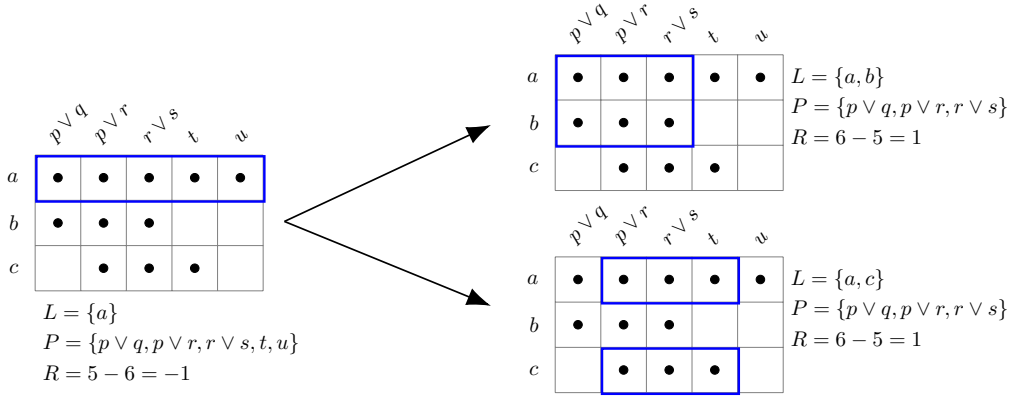

The reason for the BVA's sensitivity to randomization is due to a detail in the way implementations treats ties between literals. As described in \autoref{sec:preliminaries}, the algorithm chooses the literal that maximizes the number of remaining resolvents to be eliminated (\autoref{alg:bva}, line 4). If there is a tie between two literals, the original algorithm does not specify which literal should be used. The original implementation provided by \cite{manthey2012automated} breaks ties using the variable number in the original formula. \autoref{fig:bva_ties} shows how breaking ties differently leads to different variable additions. Note that since BVA eliminates the clauses in the grid when adding a variable, it is \textit{not} possible for multiple applications of BVA to eventually add both variables resulting from a tie.

 In the $D_{5, 10}$ packing problem, colors 9 and 10 are almost fully connected; coloring a tile with color 10 means that no tile within 10 spaces of it can also be colored 10. When BVA creates a variable for these pairwise constraints, all of the clauses are tied for the number of preserved resolvents (since every pair of color-10 variables appears in a at-most-one-distance clause). Since the original implementation used variable number to break ties and ordered variables from top-left to bottom-right, the variable additions it produces follow that structured pattern. However, when the variable order is randomized, the resulting region lacks structure and the formula takes longer to solve.

\subsection{Recovering Structure}
\label{sec:recovering-structure}

After randomization, BVA struggles to introduce variables that represent coherent clusters of tiles. However, we note that the original structure is still captured by the original formula as a whole. For example, in the $D_{5, 10}$ packing problem, two variables representing color 1, $v_{i,1}$ and $v_{j,1}$, only share a pairwise constraint if they are adjacent (i.e. if $i$ and $j$ represent adjacent tiles). If we could recover a generic metric for how \textit{close} variables are to each other (e.g. in the 2D space of $D_{5, 10}$), this metric could be used to help BVA recover structure in problems where the original variable order does not result in structured variable additions.

The intuition for our heuristic, which is detailed in \autoref{sec:heuristic}, is based on the structure observed in the packing problem. We notice that while variables in color 10 are indistinguishable after randomization (i.e. all fully connected with At-Most-One-Distance clauses), the variables in color 1 preserve the structure of the original problem: these variables only share At-Most-One-Distance clauses with their immediate neighbors. Additionally, variables for the same \textit{tile location} but different \textit{colors} are all linked by an At-Least-One-Color constraint, even after randomization. One could deduce which variables in color 10 are neighbors by looking at the connectivity of the equivalent tile positions in color 1. Specifically this requires 3 ``hops'' through clauses: starting at a variable $v_{i,10}$ in color 10, we find $v_{i,1}$ in color 1 (via an At-Least-One-Color clause), then find $v_{j,1}$ in color 1 (via an At-Most-One-Distance clause), and finally find $v_{j,10}$ in color 10 (via an At-Least-One-Color clause); the full path is $v_{i,10} \rightarrow v_{i,1} \rightarrow v_{j,1} \rightarrow v_{j,10}$.

While it is possible to construct an algorithm to recover this structure specifically for the k-coloring packing problem, we generalize this concept by \textit{counting} paths. Specifically, between two variables $v_{i,10}$ and $v_{j,10}$ in color 10 there are \textit{many} paths of length 3: for example $v_{i,10} \rightarrow v_{a,10} \rightarrow v_{b,10} \rightarrow v_{j,10}$ (using only At-Most-One-Distance clauses). However, only \textit{adjacent} variables in color 10 will have the additional path that goes through color 1: $v_{i,10} \rightarrow v_{i,1} \rightarrow v_{j,1} \rightarrow v_{j,10}$. For a given variable in color 10, it will have the most 3-hop paths to variables of the immediately adjacent grid tiles. We formalize this intuition in \autoref{sec:heuristic}.

\section{Structured Reencoding}
\label{sec:heuristic}

In this section we define our implementation of a heuristic for breaking ties during variable selection in BVA. While our heuristic was initially designed to mitigate the detrimental effects of randomization on the packing coloring problems, we found that it is also effective for other problems, even ones which have not been randomized. In \autoref{sec:evaluation} we show that our heuristic-guided BVA is effective on a wide variety of problems and offers a significant improvement to solve time for certain families of formulas.

\medskip
\noindent\textbf{\textsf{The 3-Hop Heuristic.}} Our heuristic is based on the intuition that BVA should prefer to break ties by adding variables that are \textit{close} to one another. In \autoref{sec:recovering-structure}, we noticed that in the k-coloring problem, there are some paths between variables that are only present when variables are \textit{close} in the problem's 2-D space. The \textit{variable incidence graph} compactly captures this notion of variable adjacency. Here we formally define a heuristic for ``variable distance'' based on the number of \textit{paths} between pairs of variables in the \textit{variable incidence graph}.

\begin{definition}\label{def:VIG}
    The \textit{Variable Incidence Graph} (VIG) of a formula $F$ is an undirected graph $G = (V, E)$ where $V$ is the set of variables in $F$, and $E$ contains an edge between variables if they appear in a clause together. The weight on an edge $(v_1, v_2)$ is the number of clauses in which $v_1$ and $v_2$ appear together: $w(v_1, v_2) = \left| \{C \in F : \{v_1, v_2\} \subset \mathrm{Vars}(C) \} \right|$
\end{definition}

We measure variable distance by counting the number of distinct paths between two variables (i.e. using different intermediate variables or clauses). Edges in the VIG indicate the number of clauses shared by pairs of variables. For a given sequence of variables $(v_1, v_2, ..., v_n)$ the number of distinct paths through different combinations of clauses is given by $w(v_1, v_2) \cdot w(v_2, v_3) \cdot \ldots \cdot w(v_{n-1}, v_n)$. Since edge weights are multiplicative along a path, the number of different paths of length $n$ through the VIG is given by $A^n$, where $A$ is the adjacency matrix of the VIG. Since we identified that adjacent tiles in the packing problem have more length-3 paths between them, we define a simple heuristic that counts the number of paths of length 3 in the VIG, which we call the \emph{3-hop heuristic}.

\begin{definition}\label{def-3hop}

    The \textit{3-hop heuristic} $H(x, y)$ is defined as the number of distinct paths of length 3 between two variables $x$ and $y$ in the VIG. Two paths are distinct if they travel through a different sequence of variables or clauses. Given the VIG adjacency matrix $A$, the 3-hop heuristic can be computed as $H(x, y) = (A^3)_{x,y}$.
\end{definition}

We modify \autoref{alg:bva} to use our heuristic as a tie-breaker, specifically augmenting the computation of $\argmax$ in line 4: when multiple values of $l_m$ have the same number of remaining resolvents, we choose the literal $l_m$ with the highest value of $H(l, l_m)$. Our implementation of BVA, called \hbva, is written in C++ and uses the \texttt{Eigen} library for sparse matrix operations. It is capable of generating DRAT proofs describing the sequence of variable additions and clause deletions and thus could be used with a solver to generate certificates of unsatisfiability.

\begin{figure}
    \centering
    \includegraphics[width=0.45\textwidth]{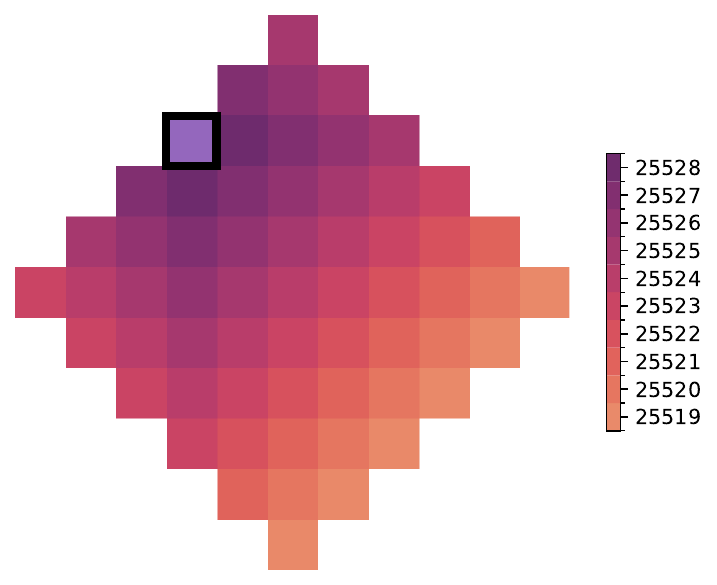}
    \caption{The value of the 3-hop heuristic in $D_{5, 10}$ between the color-10 variable for the location outlined in black and all other color-10 locations.}
    \label{fig:fig-heuristic-heatmap}
    \label{fig:fig-huuristic-packing}
\end{figure}

In \autoref{fig:fig-heuristic-heatmap}, we show the value of $H(x, y)$ in $D_{5, 10}$ for variables representing color 10 between a variable of interest (outlined in black) and all other variables of color 10. Grid tiles that are closer in the 2-D space of the packing $k$-coloring problem have more 3-hop paths between them and thus have a higher heuristic value. Using our heuristic on a randomized formula for the packing problem, we recover variables that capture the spatial structure of the problem. In the third row of \autoref{fig:first-four}, we show the first 4 variables added by \hbva, which cluster variables together using the notion of distance that is inherent in the original problem. Furthermore, we find that applying this heuristic to the packing problem results in formulas that solve much faster than BVA on a randomized formula (\autoref{table:table-packing}).

\section{Experimental Details}
\label{sec:evaluation}

We evaluated BVA on more than 29,000 formulas from the Global Benchmark Database \cite{sat_benchmark} in order to study the effects of randomization and our heuristic on BVA. In this section, we discuss the experimental setup and provide a brief overview of the results. In \autoref{sec:sec-analysis}, we analyze the results in more detail and discuss families of formulas that were significantly impacted by BVA and/or \hbva.

\medskip
\noindent\textbf{\textsf{Configurations.}} We constructed three solver configurations that use BVA in different ways. All three variants take a formula, (optionally) randomize it with \texttt{scranfilize}, run BVA (with or without heuristic), and pass it to CaDiCaL to solve. For comparison, we include a baseline variant that does not run BVA.
Since the particular ordering of clauses and variables in a formula can impact solver performance \cite{biere2019effect}, we also use the \texttt{scranfilize} tool immediately prior to running CaDiCaL in all configurations. To mitigate this variance, we run the entire sequence three times for each configuration, averaging across the three runs. The list of configurations is shown in \autoref{tab:tab-variants}. Note that all four configurations have randomization applied prior to solving with CaDiCaL but only \vrand and \vheur have randomization applied prior to BVA/\hbva.

\begin{table}[h]
    \caption{Experimental configurations. \textit{Pre} and \textit{Post} refer to arguments passed to \texttt{scranfilize} before and after running the preprocessor respectively. An empty space indicates the step was skipped for this variant.}
    \centering
    \begin{tabular}{lcccc}
        \toprule
        Variant & Pre & Preprocessor & Post & Solver \\
        \midrule
        \vbase &  &  & \texttt{-p -P -f 0.5} & CaDiCaL \\
        \vnorand &  & BVA & \texttt{-p -P -f 0.5} & CaDiCaL \\
        \vrand & \texttt{-p -P -f 0.5} & BVA & \texttt{-p -P -f 0.5} & CaDiCaL \\
        \vheur & \texttt{-p -P -f 0.5} & \hbva & \texttt{-p -P -f 0.5} & CaDiCaL \\
        \bottomrule
    \end{tabular}
    \label{tab:tab-variants}
\end{table}

\medskip
\noindent\textbf{\textsf{Benchmarks.}}
We evaluated our variants on 29\,402 benchmark instances (downloaded on February 20, 2023) from the Global Benchmark Database (GBD)~\cite{sat_benchmark}. We also report results against the Anniversary Track from the SAT Competition 2022 \cite{satcomp22} (labeled as ``ANNI-2022'' within this paper) which is included as a subset in the GBD (5355 benchmarks).

\medskip
\noindent\textbf{\textsf{Hardware.}}
All experiments were performed on the Bridges-2 system at the Pittsburgh Supercomputing Center \cite{bridges_2} on nodes with 128 cores and 256 GB RAM.

\medskip
\noindent\textbf{\textsf{Experimental Setup.}}
We compare the four configurations in a simulated competition setting with a fixed time limit of 5\,000 seconds per benchmark. The total time is computed as the sum of BVA and CaDiCaL runtimes (scranfilize time is not counted towards this limit).
As noted by Manthey et al.\ \cite{manthey2012automated}, BVA can be quite expensive, even on formulas that do not reduce significantly. We allow all versions of BVA to run for 200 seconds and if it has not terminated by then, we instead run the original formula with CaDiCaL. On our full benchmark, BVA terminates within 200 seconds on approximately 95\% of problems. 
We ran 128 instances in parallel per node, leaving approximately 2GB of memory (for reference, in the SAT Competition 2022 \cite{satcomp22}, solvers were allotted 128GB) for each BVA/CaDiCaL process. This limit is enough for most formulas, but in cases where BVA runs out of memory, we instead run the original formula in CaDiCaL. In both cases (timeout and out-of-memory), the \textit{already-used} time is added to the subsequent solve time of the original formula. This setup provides a fair comparison as BVA could be realistically configured this way in a competition.

We report the PAR-2 scores and number of formulas solved for each variant in \autoref{tab:tab-scores}. The PAR-2 score is computed as the total time it took to solve an instance (BVA runtime + CaDiCaL runtime) or twice the time limit if the formula was not solved within 5\,000 seconds. We compute the PAR-2 score individually for each run and average across the three runs of a given formula. A formula is marked as solved in \autoref{tab:tab-scores} if any of the three runs solved it within the time limit. Additionally, the set of formulas over which PAR-2 is computed consists of instances where at least one of the four configurations was able to solve it. Instances that were not solved by any configuration were not included in the PAR-2 score. Adding these entirely unsolved instances would not change the number solved and would simply scale the PAR-2 scores equally for all configurations.

\begin{table}
    \caption{PAR-2 scores and number of formulas solved for each variant split by problem type (ALL/UNSAT/SAT) and dataset (FULL/ANNI-2022). Bold cells indicate the lowest PAR-2 score or highest number solved for that group.}
    \centering
   \begin{adjustbox}{max width=\textwidth}
    \begin{tabular}{cl|rr|rr|rr}
        \toprule
        & & \multicolumn{2}{c|}{ALL} & \multicolumn{2}{c|}{UNSAT} & \multicolumn{2}{c}{SAT} \\
        Dataset & Variant & PAR-2 & \# & PAR-2 & \# & PAR-2 & \# \\
        \midrule
        \multirow{4}{*}{FULL} & \vbase & 1077.91 & 21602 & 756.14 & 6495 & 1196.99 & 15107 \\
        & \vnorand & 867.04 & 22140 & \textbf{635.71} & 6562 & 948.85 & 15578 \\
        & \vrand & 870.20 & 22077 & 673.58 & 6533 & 953.25 & 15544 \\
        & \vheur & \textbf{862.29} & \textbf{22173} & 650.41 & \textbf{6568} & \textbf{935.38} & \textbf{15605} \\
        \midrule
        \multirow{4}{*}{ANNI-2022} & \vbase & 1262.18 & 3953 & 1164.61 & 2048 & \textbf{1309.41} & 1905 \\
        & \vnorand & \textbf{1174.80} & 3987 & \textbf{967.85} & 2085 & 1338.31 & 1902 \\
        & \vrand & 1193.27 & 3958 & 1053.75 & 2060 & 1350.09 & 1898 \\
        & \vheur & 1188.63 & \textbf{3995} & 982.84 & \textbf{2088} & 1350.98 & \textbf{1907} \\
        \bottomrule
    \end{tabular}
    \end{adjustbox}
    \label{tab:tab-scores}
\end{table}

\section{Results and Analysis}
\label{sec:sec-analysis}

This section takes a closer look at the performance of \vnorand, \vrand, and \vheur in comparison to the \vbase configuration. We explore both the effects of randomization and the effects of the heuristic, in general and on specific families of formulas. 
Specifically, we explore the following questions:

\begin{enumerate}
    \item Does compression factor correlate with solve time in the context of BVA?
    \item What is the effect of randomization on the performance of BVA?
    \item Can our heuristic outperform randomized BVA?
    \item How does the performance of our heuristic vary across different families of formulas?
\end{enumerate}

We address these questions directly in the following paragraphs:

\medskip
\noindent\textbf{\textsf{A1: Formulas with larger compression factors tend to be solved faster, but this is not always the case.}}
As demonstrated in \autoref{table:table-packing}, even small reductions from BVA can have a large impact on solve time. For example, on the packing $k$-coloring problem, a \textit{single added variable} can reduce solve time by over a factor of $5$ if picked correctly (\autoref{table:table-packing}).

\begin{figure}
    \centering
    \includegraphics[width=0.8\textwidth]{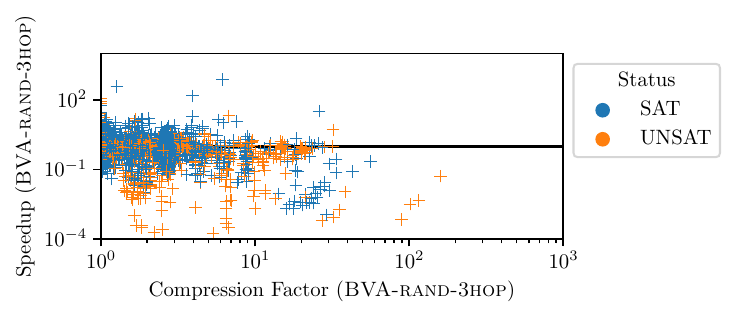}
    \caption{Formula speedup compared to compression factor for \vheur.}
    \label{fig:fig-reduction-speedup-heuristic}
\end{figure}

We compute the \textit{compression factor} of a formula as the ratio of the formula size \textit{before} to the size \textit{after} running BVA. For example, a factor of 1 indicates no reduction, a factor of 2 indicates the formula was reduced to 50\% the original size, and a factor of 10 indicates the formula was reduced to 10\% of the original size. Similarly, we compute the \textit{speedup} as the ratio of solve time to \vbase solve time (values below 1 indicate the formula was solved faster). In \autoref{fig:fig-reduction-speedup-heuristic} we plot the speedup of \vheur against the compression factor for every problem in the benchmark. Equivalent figures for \vnorand and \vrand look similar and are available in the appendix (\autoref{fig:fig-reduction-speedup-no-rand} and \autoref{fig:fig-reduction-speedup-rand}).

For formulas that could be greatly reduced, there is an observable trend towards a greater speedup. However, for small reductions, the speedup is much more variable. In some cases, even formulas that are reduced to less than 10\% of the original size may be \textit{slowed down} by BVA.
With \vheur, 60\% of formulas had a compression factor greater than 1, 40\% had a factor greater than 2, and 4\% had a factor greater than 10.

\medskip
\noindent\textbf{\textsf{A2: Randomization is Detrimental to BVA.}}
Randomization has a negative effect on the performance of BVA; in all benchmark groups, \vrand solved fewer formulas and has a higher PAR-2 score than \vnorand. Interestingly, this effect appears to be entirely due to the \textit{structure} of the resulting formula and not the resulting \textit{size} of the formula.

\begin{figure}
    \centering
    \captionsetup[subfigure]{justification=centering}
    \begin{subfigure}[t]{0.4\textwidth}
        \centering
        \includegraphics[width=\textwidth]{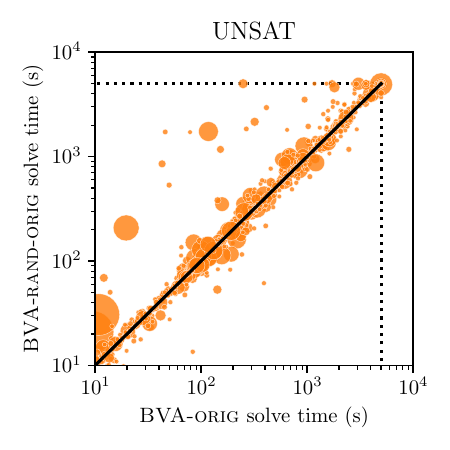}
        \caption{Relative solve time for UNSAT formulas.}
        \label{fig:fig-rand-unsat}
    \end{subfigure}
    \begin{subfigure}[t]{0.4\textwidth}
        \centering
        \includegraphics[width=\textwidth]{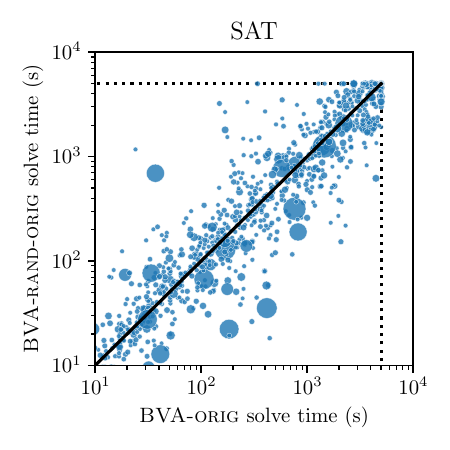}
        \caption{Relative solve time for SAT formulas.}
        \label{fig:fig-rand-sat}
    \end{subfigure}
    \caption{Difference in reduction size and solve time between \vrand and \vnorand on formulas from ANNI-2022. Larger points indicate a more-reduced formula.}
    \label{fig:fig-effect-rand}
\end{figure}

In \autoref{fig:fig-effect-rand}, we plot the relative solve times of formulas from the ANNI-2022 benchmark for \vrand and \vnorand. While there is almost no difference in the reduction sizes of the formulas produced by \vrand and \vnorand (formula sizes differ by less than 1.5\%), a number of formulas were substantially slowed down (\autoref{fig:fig-effect-rand}). Note that in these plots, randomization prior to BVA is more detrimental for UNSAT formulas and introduces a lot of variance to SAT formulas.

\medskip
\noindent\textbf{\textsf{A3: 3-Hop Heuristic is Robust to Randomization.}}
While randomization has a negative effect on the original implementation of BVA, we observe that our heuristic-guided BVA is robust to this effect. Despite being provided with randomized formulas, it is able to generate high quality variable additions and recover all of the performance loss of \vrand, even surpassing \vnorand in many cases on number of problems solved and PAR-2 score. We believe the slight performance improvement over \vnorand in several cases is due to the presence of ``pre-randomized'' formulas in the benchmark; in these cases \vnorand already suffers the effects of randomization while \vheur is able to recover the original structure of the problem.

\begin{figure}[t]
    \centering
    \captionsetup[subfigure]{justification=centering}
    \begin{subfigure}[t]{0.4\textwidth}
        \centering
        \includegraphics[width=\textwidth]{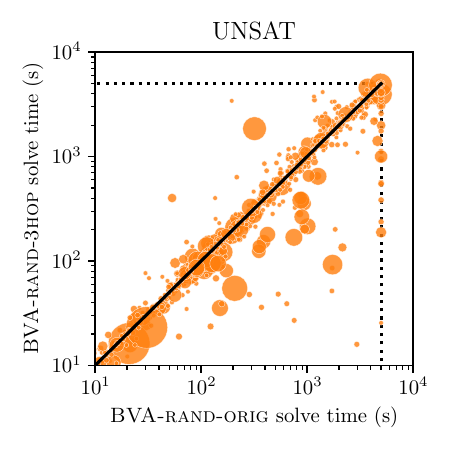}
        \caption{Relative solve time for UNSAT formulas.}
        \label{fig:fig-heur-unsat}
    \end{subfigure}
    \begin{subfigure}[t]{0.4\textwidth}
        \centering
        \includegraphics[width=\textwidth]{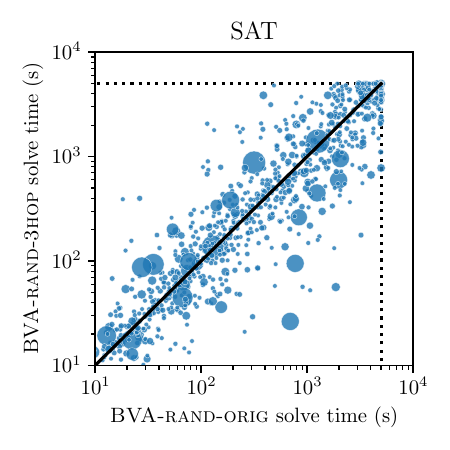}
        \caption{Relative solve time for SAT formulas.}
        \label{fig:fig-heur-sat}
    \end{subfigure}
    \caption{Difference in reduction size and solve time between \vheur and \vrand on formulas from ANNI-2022. Larger points indicate a more-reduced formula.}
    \label{fig:fig-effect-heur}
\end{figure}

In \autoref{fig:fig-effect-heur}, we compare the relative solve times of formulas from the ANNI-2022 benchmark for \vheur and \vrand. As in the previous section, the formula sizes between the two variants differs by less than 1.5\% on average. However, \vheur is able to speed up many formulas, especially UNSAT instances.

\begin{figure}[h]
    \centering
    \includegraphics[width=\textwidth]{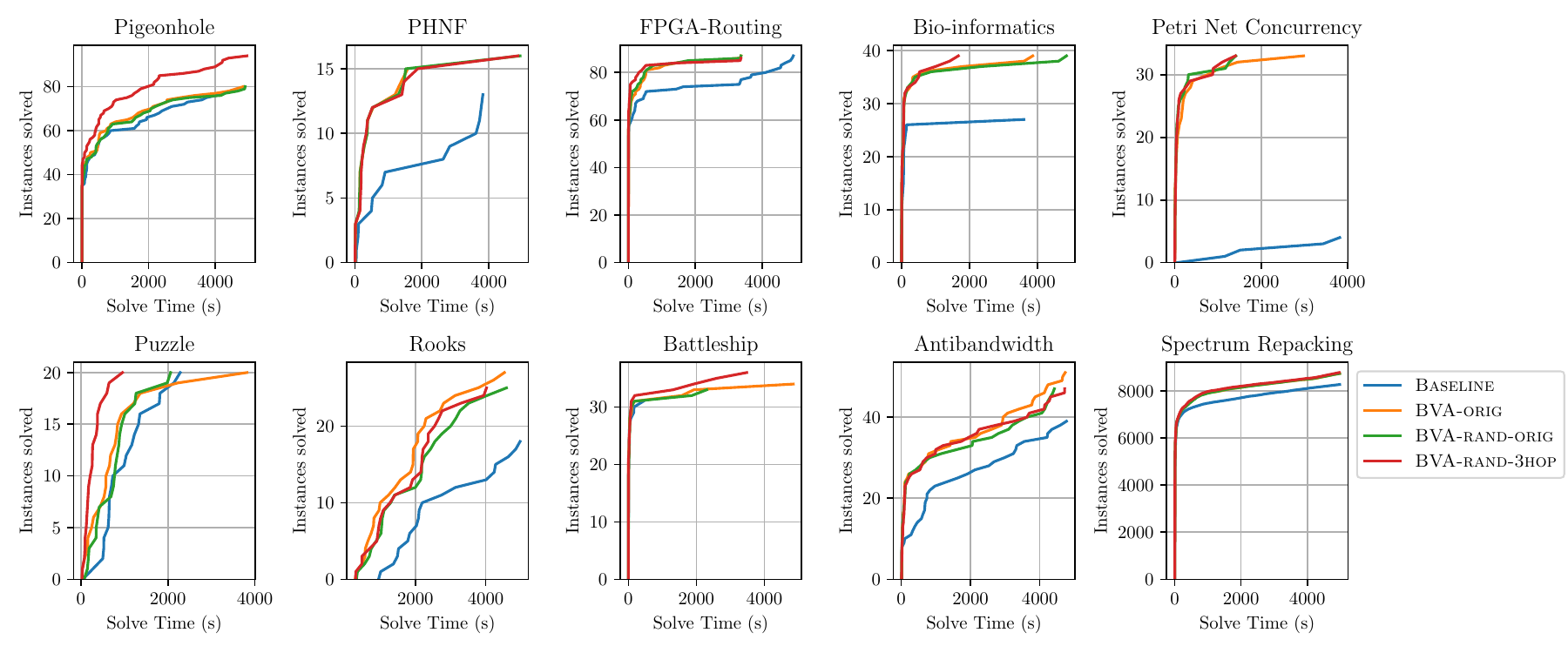}
    \caption{Performance of BVA/\hbva on 10 families of formulas where it was effective.}
    \label{fig:fig-effective}
\end{figure}

\medskip
\noindent\textbf{\textsf{A4: \hbva performs similar to BVA in most cases and performs extremely well for a few families.}}
We found that both the original implementation of BVA and our heuristic-guided version have strong effects for specific families of formulas. In \autoref{fig:fig-effective}, we plot the relative performance of the four configurations on 10 formula families for which BVA was effective. For these plots we allow BVA/\hbva to run for the full 5\,000 seconds and consider only the CaDiCaL solve time in the plots in order to understand the effectiveness of the formula rather than the speed of BVA. In this section, we briefly describe some of the families where BVA was most effective.

\medskip
\noindent\textbf{\textsf{Pigeonhole / PHNF / FPGA-Routing.}}
Pigeonhole formulas try to uniquely assign $n$ pigeons to $m$ holes. Like the packing k-coloring problem, these formulas consist primarily of AtLeastOne constraints (a pigeon must be in at least one hole) and pairwise AtMostOne constraints (two pigeons cannot share a hole). Our benchmark also contains variants of this problem, e.g. allowing multiple pigeons in a hole. These formulas are difficult for SAT solvers due to the number of possible permutations.

We found that \hbva was quite effective for UNSAT instances of pigeonhole problems (note that SAT instances of pigeonhole problems are trivial), able to solve new instances that the other three configurations could not solve. Interestingly, we found that these newly solved problems consist mainly of \textit{pre-shuffled} pigeonhole problems. A full list of solved UNSAT pigeonhole problems is provided in \autoref{tab:tab-pigeonhole}.
Other pigeonhole-like families in the dataset include PHNF (Pigeonhole Normal Form)~\cite{SAT2CSP} and FPGA-Routing~\cite{nam2001comparative}, which consists of problems generated by combining two pigeonhole problems. All forms of BVA were very effective on these problems compared to \vbase.

\medskip
\noindent\textbf{\textsf{Petri Net Concurrency.}}
Petri nets are a model of concurrent computation that consists of places and transitions~\cite{petri}. They are used to model a variety of systems, including chemical reactions, manufacturing processes, and computer programs. The Petri Net Concurrency family consists of formulas that encode the satisfiability of Petri nets. All three configurations of BVA are able to generate very compact encodings for these formulas, with an average compression factor of more than 20.

\medskip
\noindent\textbf{\textsf{Bioinformatics.}}
The bioinformatics family consists of problems that encode genetic evolutionary tree computations into SAT~\cite{bonet2009efficiently}. As noted by the authors of the original BVA paper, these problems are also reduced significantly with BVA. For the problems in this family, we found that the average compression factor was more than 7 for all three BVA configurations, i.e. the formulas were reduced to less than 15\% total size on average.

\medskip
\noindent\textbf{\textsf{Puzzle / Rooks / Battleship.}}
We found BVA to be useful in several families of formulas derived from 2-D games. The puzzle family consists of formulas that encode the satisfiability of a sliding-block puzzle and were contributed by van der Grinten to SAT Comp 2017. The rooks family asks if it is possible to place $N+1$ rooks on a $N \times N$ chessboard such that no two rooks can attack each other \cite{manthey2014too}. The battleship family consists of problems that are derived from the battleship guessing game and were contributed by Skvortsov to SAT Comp 2011. BVA was effective in all three families and \hbva was especially effective for the puzzle and battleship families.

\medskip
\noindent\textbf{\textsf{Antibandwidth / Spectrum Repacking.}}
The antibandwidth \cite{fazekas2020duplex} and spectrum repacking \cite{newman2017deep} formulas are both related to assigning radio stations to channels. Specifically, the antibandwidth family asks if it possible to assign a given set of stations to a given set of channels such that the difference in channel between any two stations is at least $k$. Similarly, the spectrum repacking family asks if it is possible to reassign a given set of stations into a smaller set of channels, taking into account physical distances between stations and the bandwidth of each channel. All configurations of BVA were effective on these problems.

\section{Conclusion}

Bounded Variable Addition is surprisingly effective at reducing the size of formulas and improving solve time by introducing auxiliary variables. We discovered that this speedup is caused not only by the reduction in formula size but also the introduction of certain \textit{effective} auxiliary variables. We found that the original implementation was sensitive to randomization and proposed a new heuristic-guided implementation, \hbva, that is robust to this effect.
In a competition-style benchmark, we show that using SBVA resulted in the most formulas solved in every category, outperforming both BVA and the baseline (no preprocessor). Additionally, SBVA was extremely effective on certain families of formulas, demonstrating that auxiliary variables can be useful in practice if they are chosen carefully.

\bibliography{paper}

\newpage

\appendix
\markboth{APPENDIX}{}

\setcounter{table}{0}
\renewcommand{\thetable}{A\arabic{table}}

\setcounter{figure}{0}
\renewcommand{\thefigure}{A\arabic{figure}}

\section{Appendix}

\subsection{Reduction Size vs. Solve Time}

\begin{figure}[h]
    \centering
    \includegraphics[width=0.8\textwidth]{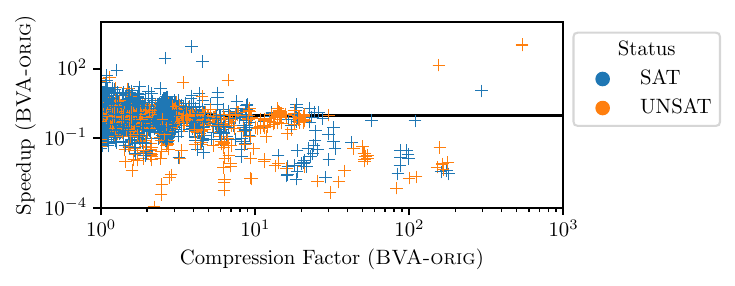}
    \caption{Formula speedup compared to compression factor for \vnorand.}
    \label{fig:fig-reduction-speedup-no-rand}
\end{figure}

\begin{figure}[h]
    \centering
    \includegraphics[width=0.8\textwidth]{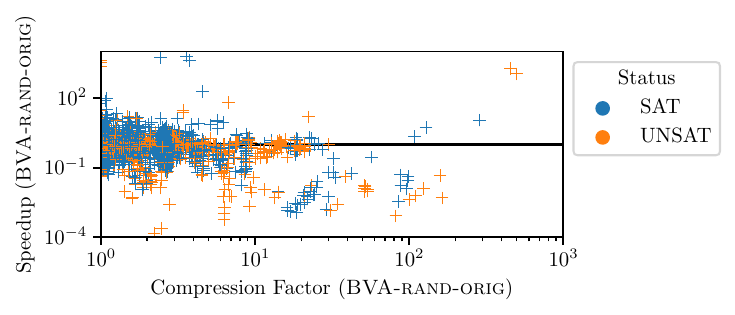}
    \caption{Formula speedup compared to compression factor for \vrand.}
    \label{fig:fig-reduction-speedup-rand}
\end{figure}

\subsection{Performance on Pigeonhole Problems}

\begin{table}
    \centering
    \caption{Performance on unsatisfiable instances of problems in the pigeon-hole family.}
    \label{tab:tab-pigeonhole}
    \resizebox*{\textwidth}{!}{\begin{tabular}{lrrrr}
        \toprule
        & \multicolumn{4}{c}{Solve Time (s)} \\
        Instance (unsatisfiable) & \vbase & \vnorand & \vrand & \vheur \\
        \midrule
        a\_rphp035\_05 & 499.77 & 478.28 & \textbf{327.31} & 328.82 \\
        a\_rphp045\_05 & 2165.96 & 2069.67 & \textbf{1748.25} & 1952.39 \\
        a\_rphp055\_04 & 79.75 & 81.91 & 79.09 & \textbf{75.39} \\
        a\_rphp065\_04 & 168.98 & 164.66 & \textbf{137.85} & 146.82 \\
        a\_rphp085\_04 & 760.00 & 752.18 & 707.73 & \textbf{661.66} \\
        a\_rphp098\_04 & \textbf{1945.51} & 2136.84 & 2726.63 & 2318.10 \\
        \midrule
        ae\_rphp035\_05 & 410.54 & 501.24 & 426.94 & \textbf{407.24} \\
        ae\_rphp045\_05 & 2402.73 & 2549.29 & \textbf{2206.80} & 2307.20 \\
        ae\_rphp055\_04 & 83.73 & 86.48 & 81.64 & \textbf{74.17} \\
        ae\_rphp075\_04 & 513.63 & 515.97 & 558.53 & \textbf{504.86} \\
        ae\_rphp095\_04-sc2018 & 1937.08 & 1686.12 & 2074.11 & \textbf{1678.99} \\
        ae\_rphp095\_04 & 1706.39 & 1611.31 & \textbf{1560.96} & 1572.97 \\
        \midrule
        clqcolor-08-06-07.shuffled-as.sat05-1257 & 3.28 & \textbf{0.62} & 0.91 & 0.81 \\
        counting-clqcolor-unsat-set-b-clqcolor-08-06-07.sat05-1257.reshuffled-07 & 2.69 & 0.83 & \textbf{0.76} & 1.02 \\
        \midrule
        counting-easier-fphp-012-010.sat05-1214.reshuffled-07 & 111.76 & 33.26 & 30.03 & \textbf{0.11} \\
        counting-easier-fphp-014-012.sat05-1215.reshuffled-07 & \tim & \tim & \tim & \textbf{1.62} \\
        counting-easier-php-012-010.sat05-1172.reshuffled-07 & 139.59 & 27.07 & 29.35 & \textbf{3.45} \\
        counting-easier-php-018-014.sat05-1175.reshuffled-07 & \tim & \tim & \tim & \textbf{4224.30} \\
        counting-harder-php-014-013.sat05-1187.reshuffled-07 & \tim & \tim & \tim & \textbf{1760.18} \\
        \midrule
        e\_rphp035\_05-sc2018 & 424.51 & 460.54 & 461.25 & \textbf{387.80} \\
        e\_rphp035\_05 & 482.43 & 480.30 & \textbf{419.09} & 501.65 \\
        e\_rphp055\_04 & \textbf{73.62} & 81.31 & 77.03 & 78.85 \\
        e\_rphp065\_04 & 139.17 & \textbf{125.17} & 143.14 & 144.34 \\
        e\_rphp096\_04 & 1626.72 & 1830.11 & \textbf{1492.40} & 1503.88 \\
        \midrule
        easier-fphp-020-015.sat05-1218.reshuffled-07 & \tim & \tim & \tim & \textbf{4205.14} \\
        \midrule
        fphp-010-008.shuffled-as.sat05-1213 & 0.49 & 0.22 & 0.22 & \textbf{0.02} \\
        fphp-010-009.shuffled-as.sat05-1227 & 5.43 & 4.30 & 4.22 & \textbf{0.06} \\
        fphp-012-010.shuffled-as.sat05-1214 & 113.38 & 45.04 & 36.16 & \textbf{0.12} \\
        fphp-012-011.shuffled-as.sat05-1228 & 1722.86 & 1525.19 & 1844.10 & \textbf{0.68} \\
        fphp-014-012.shuffled-as.sat05-1215 & \tim & \tim & \tim & \textbf{1.95} \\
        fphp-014-013.shuffled-as.sat05-1229 & \tim & \tim & \tim & \textbf{564.50} \\
        fphp-016-013.shuffled-as.sat05-1216 & \tim & \tim & \tim & \textbf{383.94} \\
        fphp-016-015.shuffled-as.sat05-1230 & \tim & \tim & \tim & \textbf{1027.46} \\
        fphp-018-014.shuffled-as.sat05-1217 & \tim & \tim & \tim & \textbf{549.07} \\
        fphp-020-015.shuffled-as.sat05-1218 & \tim & \tim & \tim & \textbf{944.65} \\
        \midrule
        harder-fphp-016-015.sat05-1230.reshuffled-07 & \tim & \tim & \tim & \textbf{3496.64} \\
hole10.cnf.mis-98.debugged & 2.20 & \textbf{0.95} & 1.51 & 1.05 \\
        \midrule
        ph9 & 5.68 & 1.55 & 4.02 & \textbf{0.08} \\
        ph10 & 120.63 & 13.83 & 50.17 & \textbf{8.64} \\
        ph11 & 3061.40 & 35.76 & 790.89 & \textbf{26.45} \\
        \midrule
        php-010-008.shuffled-as.sat05-1171 & 0.64 & 0.15 & 0.25 & \textbf{0.03} \\
        php-010-009.shuffled-as.sat05-1185 & 6.59 & 3.03 & 3.58 & \textbf{0.07} \\
        php-012-010.shuffled-as.sat05-1172 & 138.30 & 31.52 & 23.96 & \textbf{3.02} \\
        php-012-011.shuffled-as.sat05-1186 & 2695.99 & 1006.29 & 755.79 & \textbf{26.97} \\
        php-014-012.shuffled-as.sat05-1173 & \tim & \tim & \tim & \textbf{237.47} \\
        php-016-013.shuffled-as.sat05-1174 & \tim & \tim & \tim & \textbf{3024.16} \\
        \midrule
        php11e11 & 3599.98 & \textbf{500.38} & 780.09 & 796.55 \\
        \midrule
rphp4\_065\_shuffled & 146.66 & 148.59 & \textbf{129.96} & 132.01 \\
        rphp4\_070\_shuffled & \textbf{213.59} & 249.18 & 280.29 & 227.72 \\
        rphp4\_075\_shuffled & 448.36 & 459.23 & \textbf{415.65} & 419.72 \\
        rphp4\_080\_shuffled & 532.00 & 519.81 & 516.74 & \textbf{503.20} \\
        rphp4\_085\_shuffled & 666.98 & 723.74 & 654.52 & \textbf{648.68} \\
        rphp4\_090\_shuffled & 853.68 & \textbf{850.60} & 919.62 & 890.40 \\
        rphp4\_095\_shuffled & 1571.45 & 1362.84 & 1611.21 & \textbf{1342.79} \\
        rphp4\_100\_shuffled & 2314.76 & 2545.95 & 2361.29 & \textbf{2154.53} \\
        rphp4\_105\_shuffled & 3168.06 & 2948.21 & 2520.33 & \textbf{2249.50} \\
        rphp4\_110\_shuffled & 3726.95 & 3389.56 & \textbf{3208.10} & 3665.63 \\
        rphp4\_115\_shuffled & 4155.81 & 4406.08 & 4169.75 & \textbf{4095.25} \\
        rphp4\_120\_shuffled & \tim & \textbf{4840.06} & 4883.44 & 4948.49 \\
        rphp4\_125\_shuffled & \tim & 4613.90 & 4676.45 & \textbf{3989.60} \\
        \midrule
        rphp\_p6\_r28 & \tim & \tim & \textbf{4895.39} & \tim \\
        \midrule
        tph6 & 226.78 & 12.06 & 68.97 & \textbf{0.52} \\
        tph7 & \tim & 249.13 & \tim & \textbf{0.88} \\
        tph8 & \tim & \tim & \tim & \textbf{188.30} \\
        \bottomrule
    \end{tabular}}
\end{table}

\newpage

\subsection{Performance on Bioinformatics Problems}

\begin{table}[h!]
    \centering
    \caption{Performance on unsatisfiable instances of problems in the bioinformatics family.}
    \label{tab:tab-bio-unsat}
    \resizebox{\textwidth}{!}{\begin{tabular}{lrrrr}
        \toprule
        & \multicolumn{4}{c}{Solve Time (s)} \\
        Instance (unsatisfiable) & \vbase & \vnorand & \vrand & \vheur \\
        \midrule
        ndhf\_xits\_09\_UNSAT & \tim & \textbf{9.69} & 11.10 & 10.52 \\
ndhf\_xits\_10\_UNSAT & \tim & 70.30 & 70.43 & \textbf{63.04} \\
ndhf\_xits\_11\_UNSAT & \tim & 612.65 & 869.25 & \textbf{401.50} \\
ndhf\_xits\_12\_UNSAT & \tim & \tim & \tim & \textbf{1003.12} \\
\midrule
rbcl\_xits\_06\_UNSAT & 5.27 & 0.20 & 0.22 & \textbf{0.20} \\
rbcl\_xits\_07\_UNSAT & 110.14 & 0.60 & 0.60 & \textbf{0.47} \\
rbcl\_xits\_08\_UNSAT & 3603.02 & 2.09 & 2.15 & \textbf{1.72} \\
rbcl\_xits\_09\_UNKNOWN & \tim & \textbf{8.11} & 9.55 & 10.69 \\
rbcl\_xits\_10\_UNKNOWN & \tim & 69.24 & \textbf{56.82} & 95.77 \\
rbcl\_xits\_11\_UNKNOWN-sc2009 & \tim & 311.98 & \textbf{309.87} & 505.25 \\
rbcl\_xits\_11\_UNKNOWN & \tim & \textbf{331.32} & 392.19 & 528.49 \\
rbcl\_xits\_12\_UNKNOWN & \tim & \textbf{3607.56} & 4857.58 & \tim \\
\midrule
rpoc\_xits\_07\_UNSAT & 54.86 & \textbf{0.95} & 1.03 & 0.97 \\
rpoc\_xits\_09\_UNSAT & \tim & 41.37 & 30.13 & \textbf{24.29} \\
rpoc\_xits\_10\_UNKNOWN & \tim & 237.25 & \textbf{172.23} & 182.65 \\
rpoc\_xits\_11\_UNKNOWN-sc2009 & \tim & 3863.50 & 2369.91 & \textbf{1672.29} \\
rpoc\_xits\_11\_UNKNOWN & \tim & 1813.61 & 4624.68 & \textbf{1413.96} \\
        \bottomrule
    \end{tabular}}
\end{table}

\begin{table}[h!]
    \centering
    \caption{Performance on satisfiable instances of problems in the bioinformatics family.}
    \label{tab:tab-bio-sat}
    \resizebox{\textwidth}{!}{\begin{tabular}{lrrrr}
        \toprule
        & \multicolumn{4}{c}{Solve Time (s)} \\
        Instance (unsatisfiable) & \vbase & \vnorand & \vrand & \vheur \\
        \midrule
        ndhf\_xits\_19\_UNKNOWN-sc2011 & 143.44 & 31.96 & \textbf{9.59} & 13.37 \\
ndhf\_xits\_20\_SAT & 29.44 & \textbf{2.20} & 3.32 & 3.38 \\
ndhf\_xits\_21\_SAT & 6.27 & 2.16 & \textbf{1.13} & 2.27 \\
ndhf\_xits\_22\_SAT & 3.09 & 1.11 & \textbf{0.50} & 0.86 \\
\midrule
rbcl\_xits\_14\_SAT & 1.31 & 0.47 & \textbf{0.46} & 1.91 \\
rbcl\_xits\_18\_SAT & 0.22 & 0.04 & 0.05 & \textbf{0.03} \\
rpoc\_xits\_17\_SAT & 1.22 & 0.14 & 0.11 & \textbf{0.11} \\
        \bottomrule
    \end{tabular}}
\end{table}

\subsection{Performance per Family}

Legend:
\begin{itemize}
    \item \vbase: blue
    \item \vnorand: orange
    \item \vrand: green
    \item \vheur: red
\end{itemize}

\begin{figure}[h]
    \centering
    \resizebox*{!}{\textheight}{
        \begin{subfigure}[b]{0.45\textwidth}
            \centering
            \includegraphics[width=\textwidth]{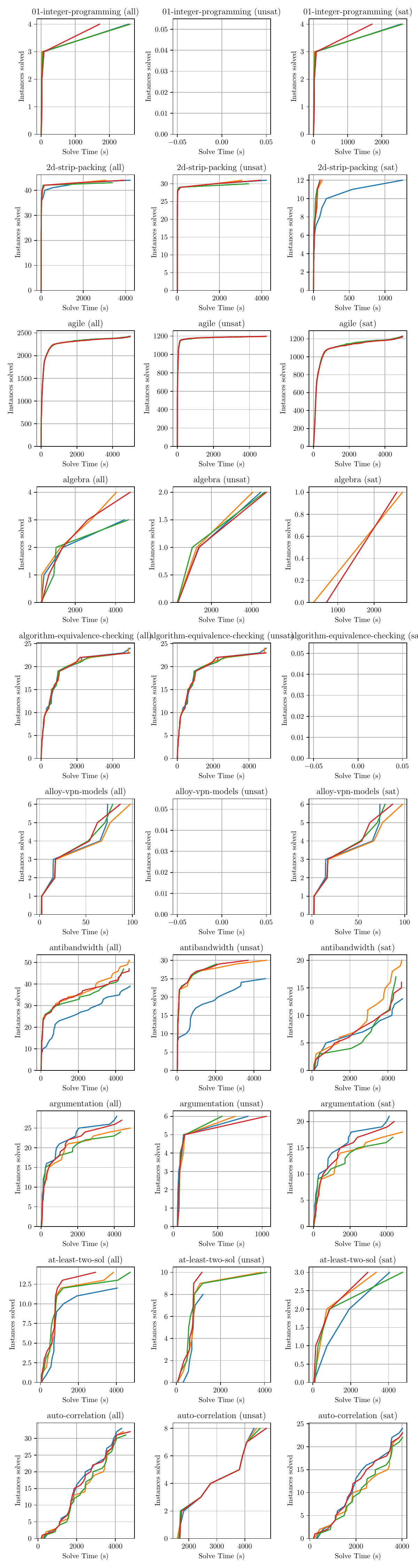}
        \end{subfigure}
        \hfill
        \begin{subfigure}[b]{0.45\textwidth}
            \centering
            \includegraphics[width=\textwidth]{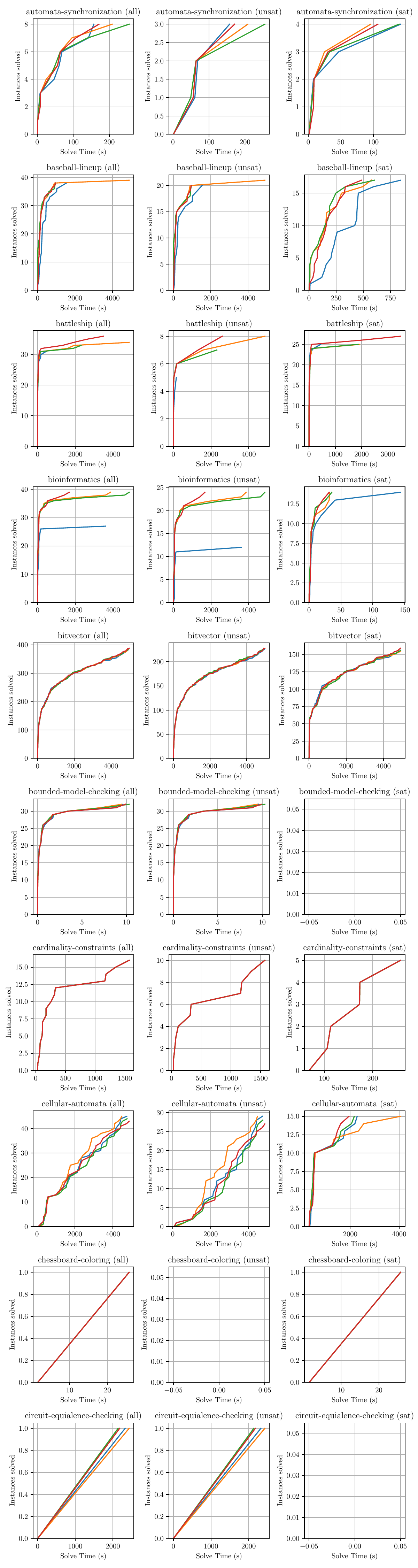}
        \end{subfigure}
    }
\end{figure}

\begin{figure}[h]
    \centering
    \resizebox*{!}{\textheight}{
        \begin{subfigure}[b]{0.45\textwidth}
            \centering
            \includegraphics[width=\textwidth]{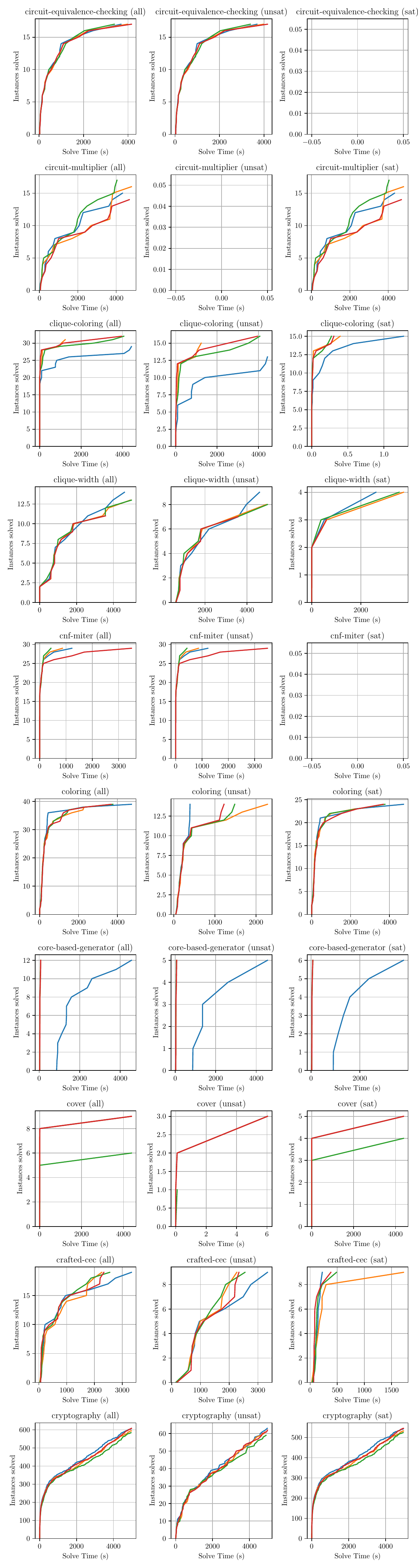}
        \end{subfigure}
        \hfill
        \begin{subfigure}[b]{0.45\textwidth}
            \centering
            \includegraphics[width=\textwidth]{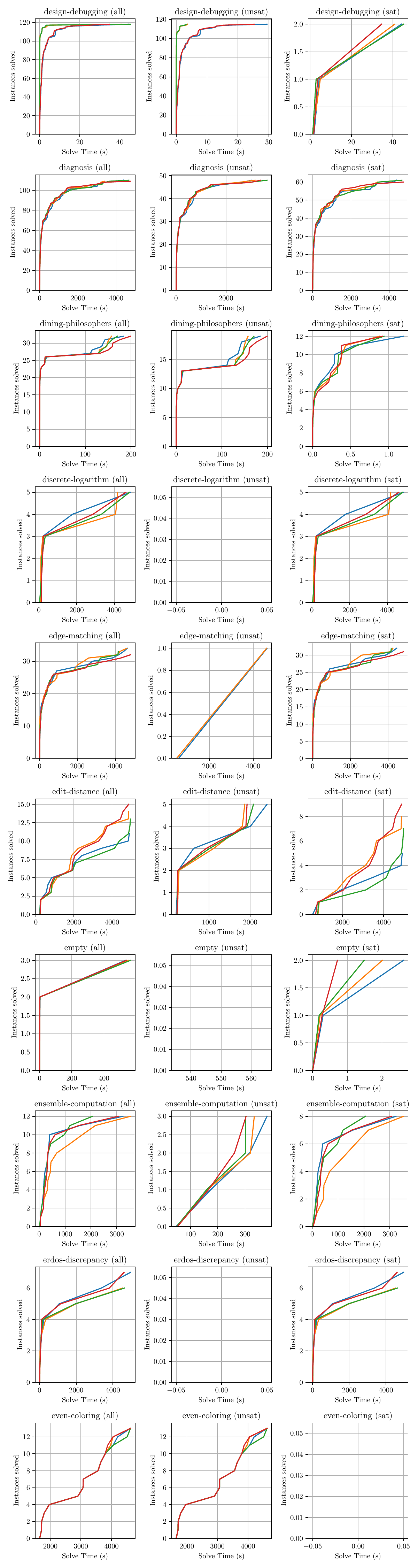}
        \end{subfigure}
    }
\end{figure}

\begin{figure}
    \centering
    \resizebox*{!}{\textheight}{
        \begin{subfigure}[b]{0.45\textwidth}
            \centering
            \includegraphics[width=\textwidth]{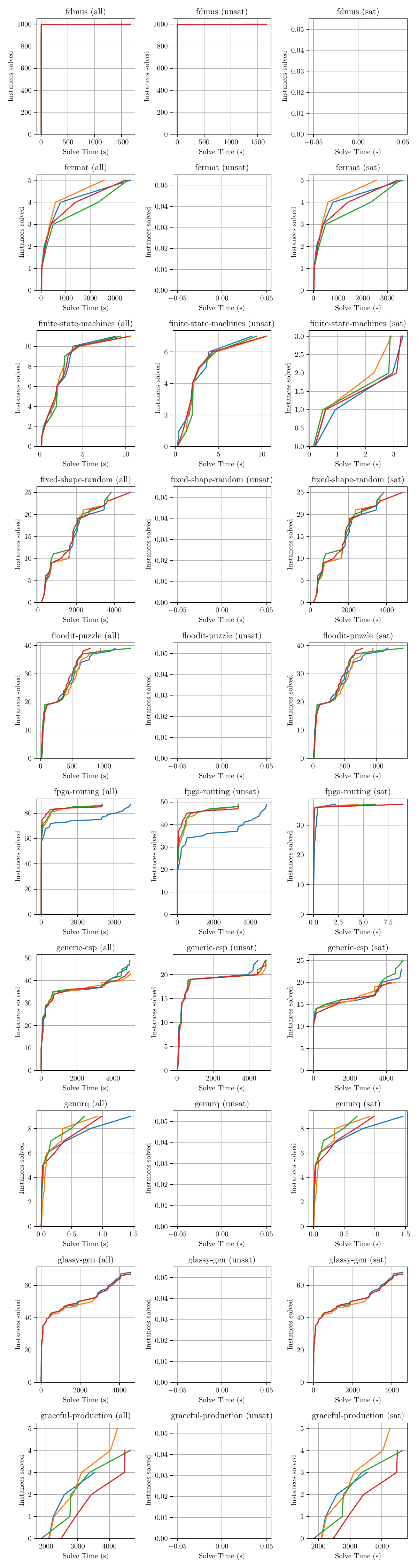}
        \end{subfigure}
        \hfill
        \begin{subfigure}[b]{0.45\textwidth}
            \centering
            \includegraphics[width=\textwidth]{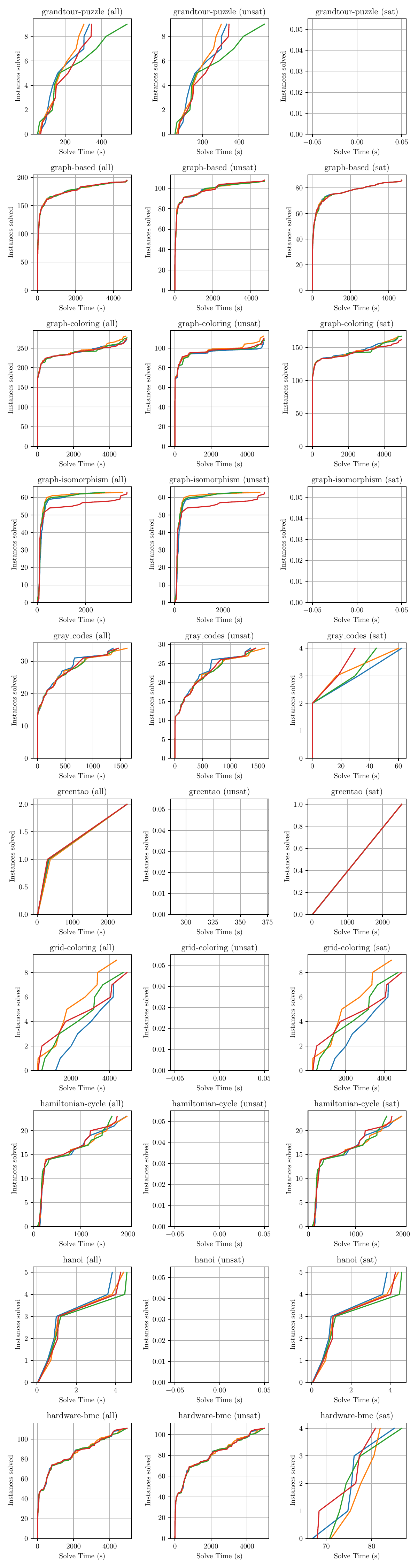}
        \end{subfigure}
    }
\end{figure}

\begin{figure}
    \centering
    \resizebox*{!}{\textheight}{
        \begin{subfigure}[b]{0.45\textwidth}
            \centering
            \includegraphics[width=\textwidth]{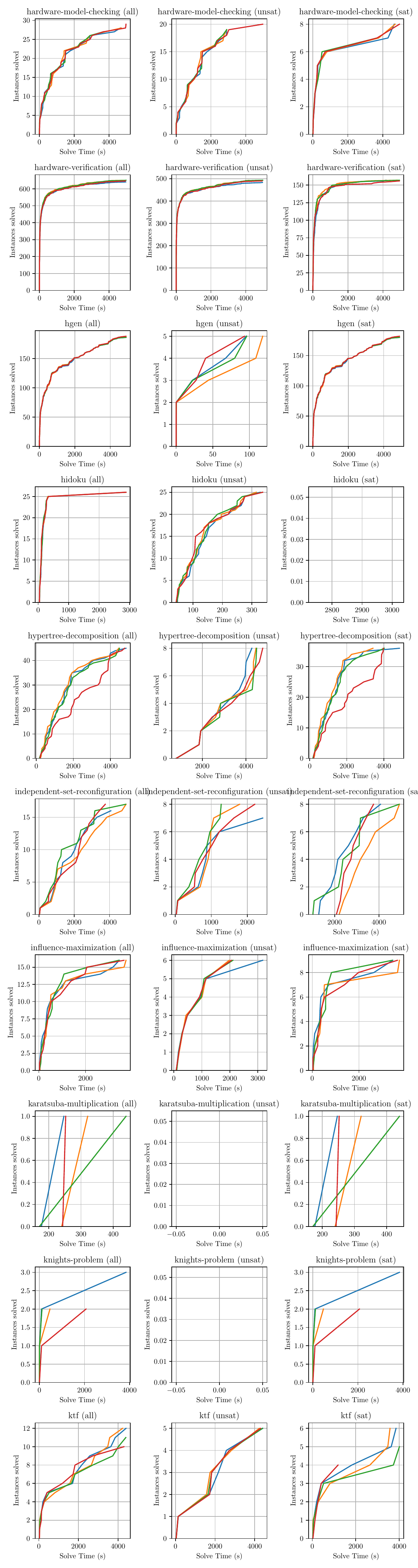}
        \end{subfigure}
        \hfill
        \begin{subfigure}[b]{0.45\textwidth}
            \centering
            \includegraphics[width=\textwidth]{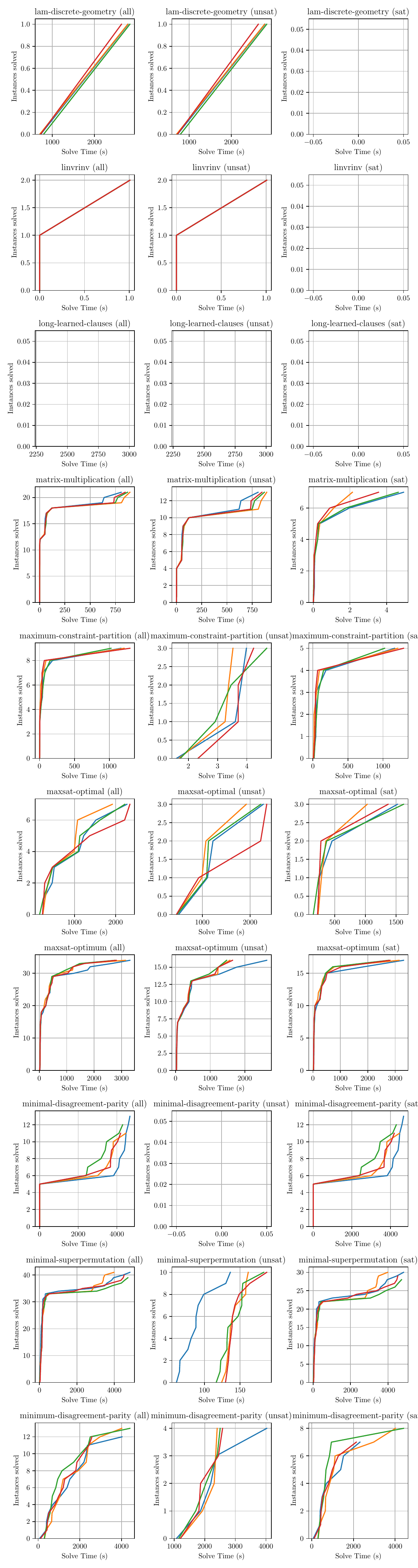}
        \end{subfigure}
    }
\end{figure}

\begin{figure}
    \centering
    \resizebox*{!}{\textheight}{
        \begin{subfigure}[b]{0.45\textwidth}
            \centering
            \includegraphics[width=\textwidth]{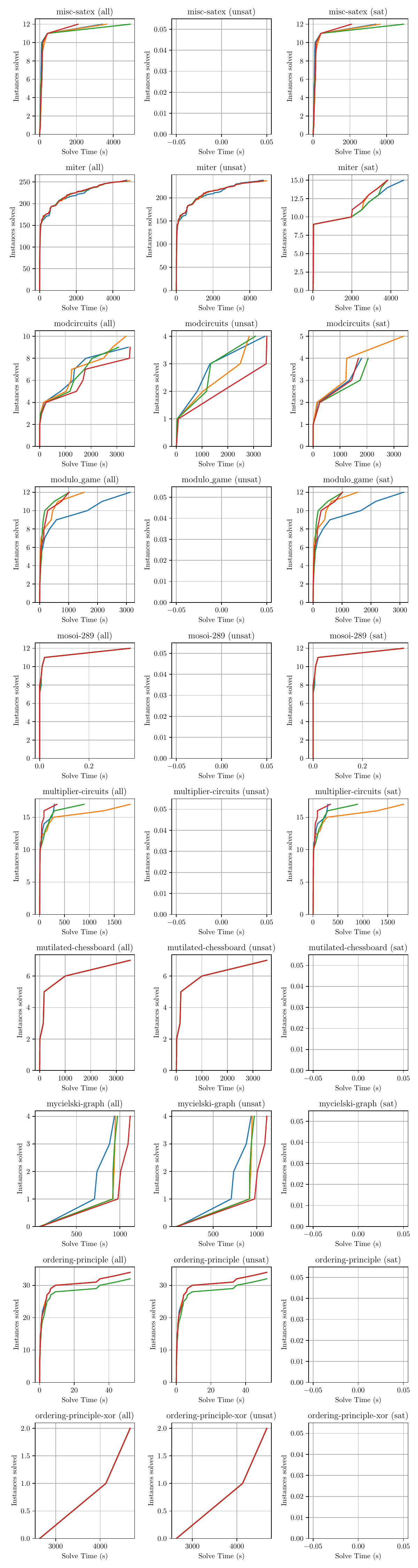}
        \end{subfigure}
        \hfill
        \begin{subfigure}[b]{0.45\textwidth}
            \centering
            \includegraphics[width=\textwidth]{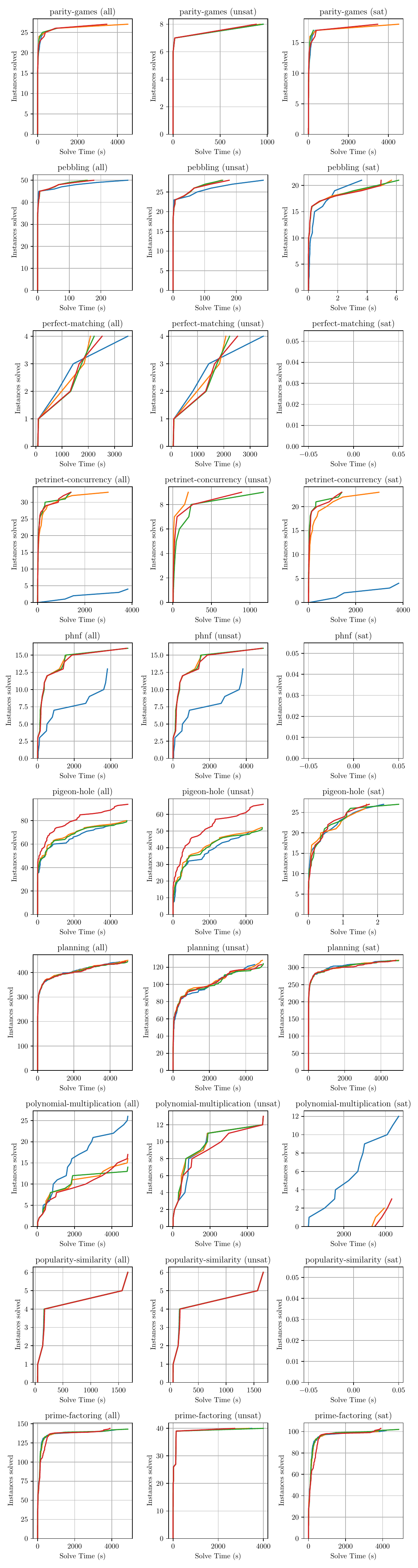}
        \end{subfigure}
    }
\end{figure}

\begin{figure}
    \centering
    \resizebox*{!}{\textheight}{
        \begin{subfigure}[b]{0.45\textwidth}
            \centering
            \includegraphics[width=\textwidth]{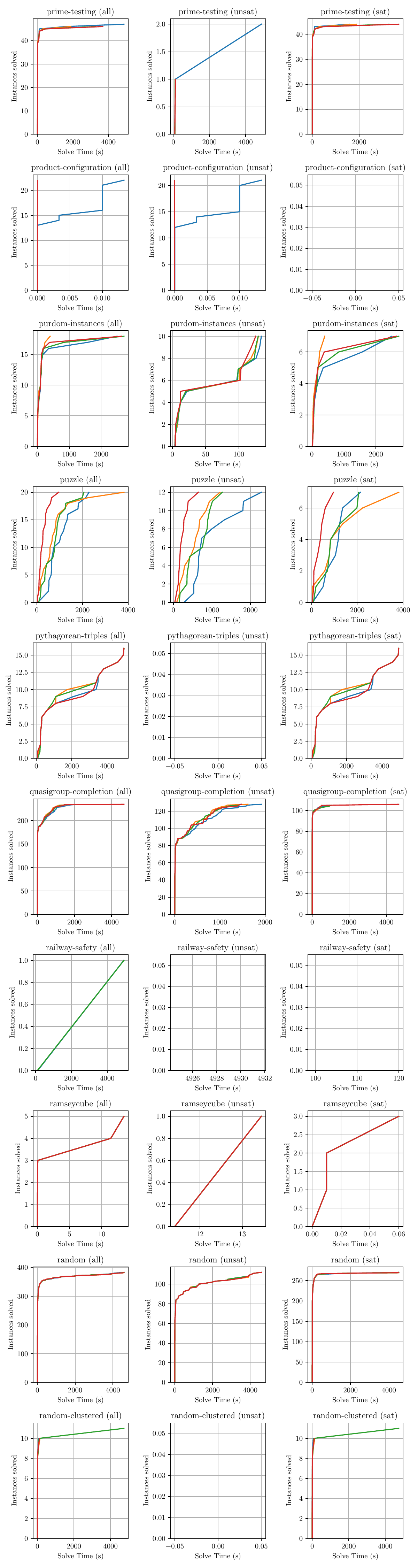}
        \end{subfigure}
        \hfill
        \begin{subfigure}[b]{0.45\textwidth}
            \centering
            \includegraphics[width=\textwidth]{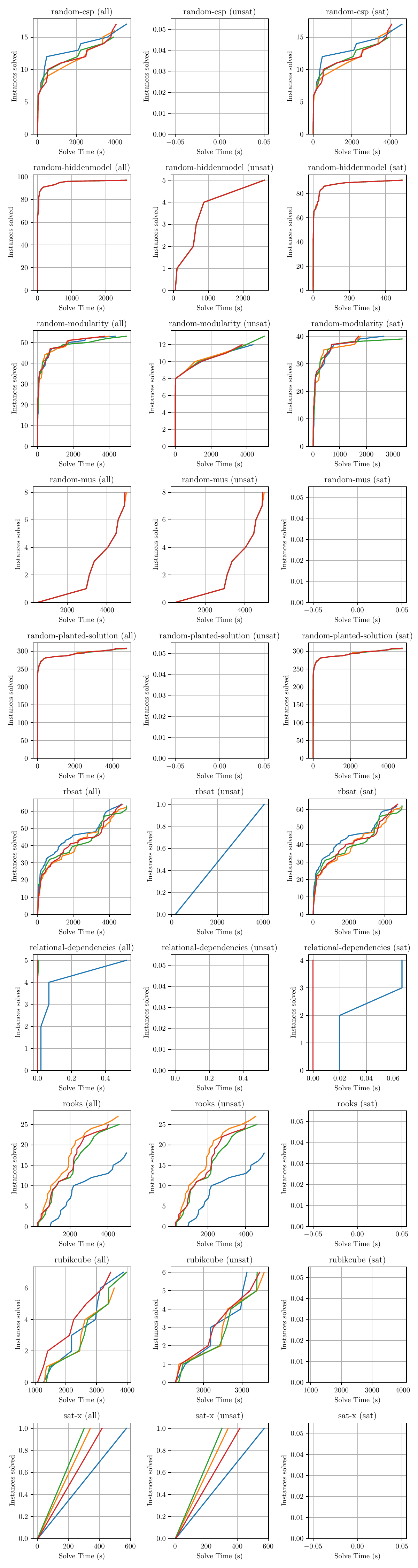}
        \end{subfigure}
    }
\end{figure}

\begin{figure}
    \centering
    \resizebox*{!}{\textheight}{
        \begin{subfigure}[b]{0.45\textwidth}
            \centering
            \includegraphics[width=\textwidth]{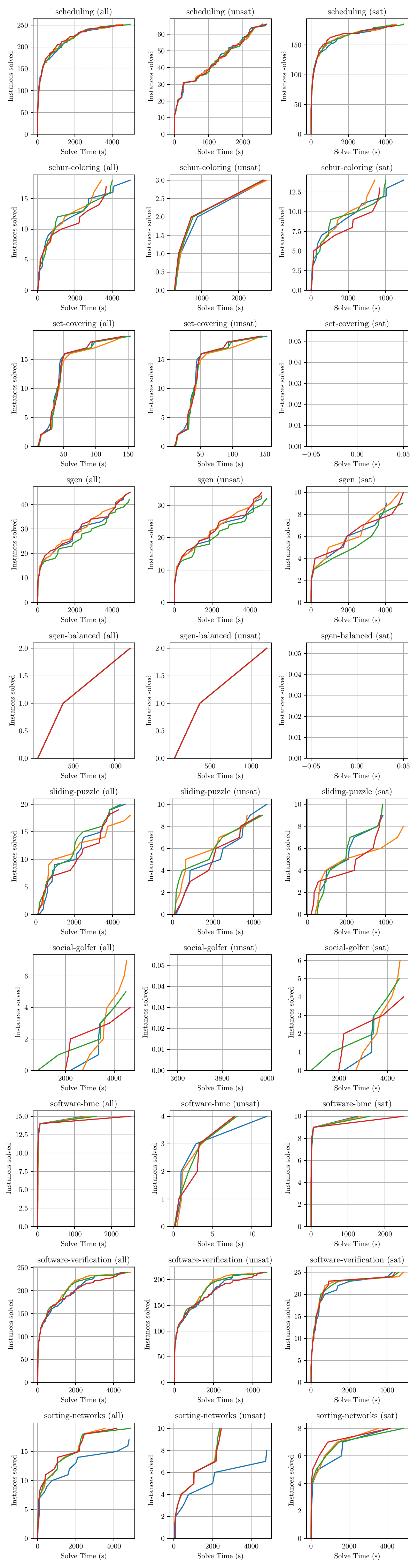}
        \end{subfigure}
        \hfill
        \begin{subfigure}[b]{0.45\textwidth}
            \centering
            \includegraphics[width=\textwidth]{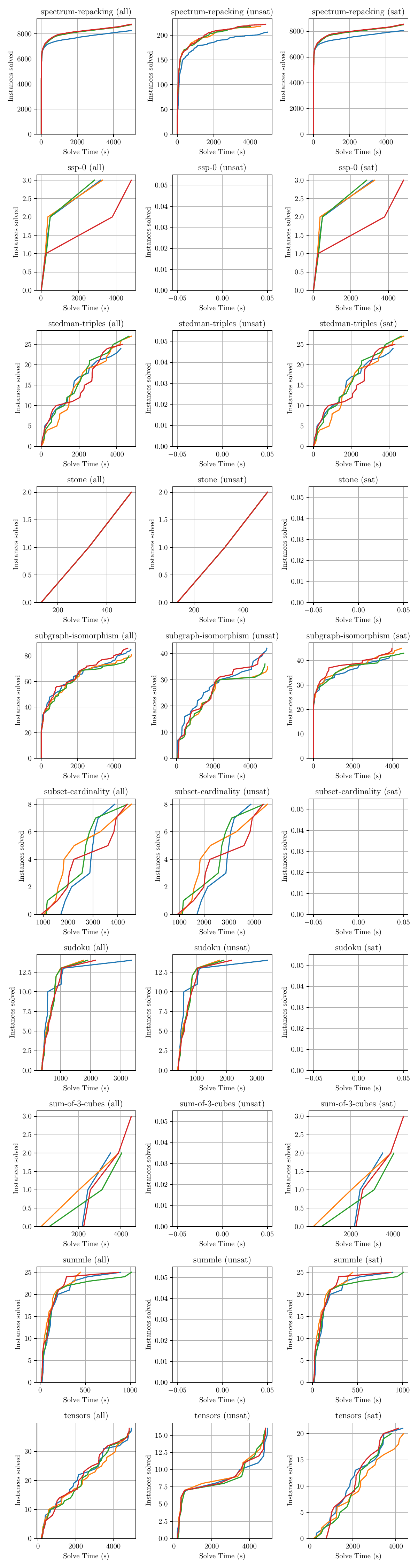}
        \end{subfigure}
    }
\end{figure}

\begin{figure}
    \centering
    \resizebox*{!}{\textheight}{
        \begin{subfigure}[b]{0.45\textwidth}
            \centering
            \includegraphics[width=\textwidth]{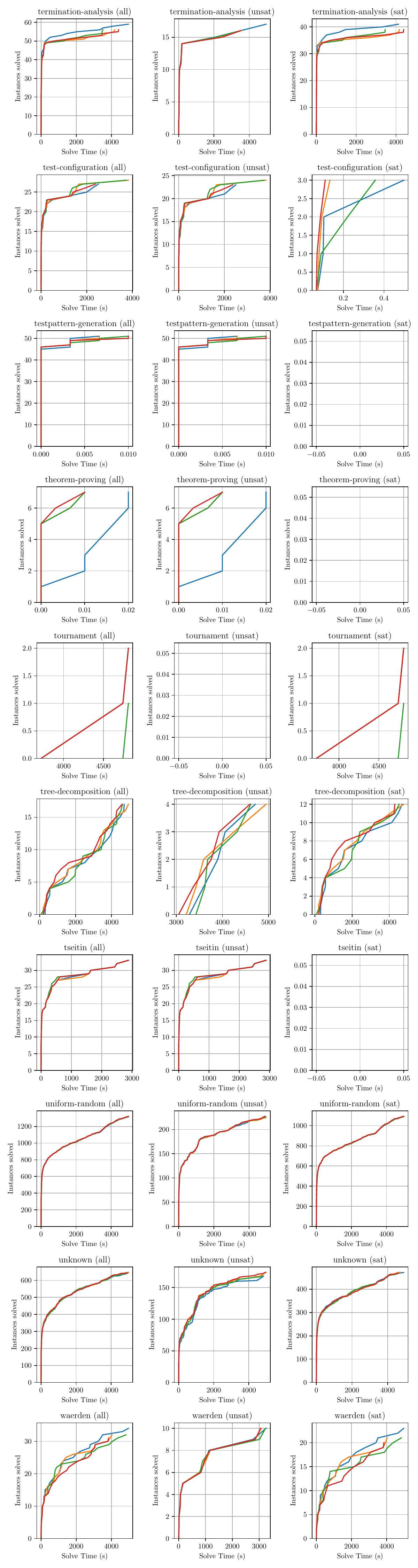}
        \end{subfigure}
    }
\end{figure}

\end{document}